\documentclass[12pt]{revtex4}

\usepackage{here}
\usepackage{graphicx}
\usepackage{epsfig}
\usepackage{amsmath}
\usepackage{amssymb}
\usepackage{color}
\usepackage{xspace}

\newcommand{\be}{\begin{displaymath}}
\newcommand{\ee}{\end{displaymath}}
\newcommand{\bn}{\begin{equation}}
\newcommand{\en}{\end{equation}}

\newcommand{\gyro}{{\sc gyro}\xspace}

\begin{document}

\title[Micro-tearing modes in spherical and conventional tokamaks]{Micro-tearing modes in spherical and conventional tokamaks}

\author{S. Moradi$^{1}$, I. Pusztai$^{1,3}$, W. Guttenfelder$^{2}$, T. F\"ul\"op$^{1}$ and A. Moll\'en$^{1}$}

\address{$^{1}$Department of Applied Physics, Nuclear Engineering, Chalmers University of Technology and Euratom-VR Association, G\"oteborg, Sweden\\
  $^{2}$Princeton Plasma Physics Laboratory, Princeton NJ 08543, USA \\
 {$^{3}$Plasma Science and Fusion Center, Massachusetts Institute of Technology, Cambridge MA, 02139, USA}}
\begin{abstract}
  The onset and characteristics of Micro-Tearing Modes (MTM) in the
  core of spherical (NSTX) and conventional tokamaks (ASDEX-UG and
  JET) are studied through local linear gyrokinetic simulations with
  {\sc gyro} [J.~Candy and E. Belli, General Atomics Report GA-A26818
  (2011)]. For experimentally relevant core plasma parameters in the
  NSTX and ASDEX-UG tokamaks, in agreement with previous works, we
  find MTMs as the dominant linear instability. Also, for JET-like
  core parameters considered in our study an MTM is found as the most
  unstable mode. In all these plasmas, finite collisionality is needed
  for MTMs to become unstable and the electron temperature gradient is
  found to be the fundamental drive. However, a significant difference
  is observed in the dependence of linear growth rate of MTMs on
  electron temperature gradient. While it varies weakly and
  non-monotonically in JET and ASDEX-UG plasmas, in NSTX it increases
  with the electron temperature gradient.
    \end{abstract}
\pacs{52.25 Fi, 52.25 Ya, 52.55 Fa}
\maketitle

\section{Introduction}
In recent years, specially in view of an increasing interest in high
$\beta_e$ operation scenarios, such as hybrid scenarios for ITER
\cite{shimada,mcdonald2008}, the impact of electromagnetic effects on
the particle and heat transport has attracted much attention. Here
$\beta_e=(8\pi n_e T_e)/B_{unit}^{2}$, where $ n_e$ and $T_e$ are the
electron density and temperature. $B_{unit}$ is defined as the
effective field strength \cite{waltz09}.

Recent reports have shown the significant role of electromagnetic
modes such as Micro-Tearing Modes (MTMs) on the electron heat
transport in the core of fusion plasmas
\cite{ApplegatePPCF2007,GuttenfelderPRL2011,DoerkPRL2011,toldPoP2008}. It
has been found that in plasmas where $\beta_e$ and collisionality are
sufficiently high MTMs can become the dominant instability. Under
these conditions, MTMs generate a major contribution to the anomalous
electron heat flux.

The MTMs are small-scale in the radial direction, but ion-scale in the
binormal direction. They are electromagnetic modes with an even parity
with respect to the perturbed parallel vector potential, $\delta
A_{\parallel}$ (referred to as \emph{tearing parity}). In the
literature two drive mechanisms for MTMs are proposed: one is the
time-dependent thermal force experienced by electrons which results in
a parallel current that produces magnetic field perturbations. If
these perturbations then tip the field lines in the direction of an
equilibrium electron temperature gradient the thermal force will
increase and therefore the instability arises \cite{drack}.  The
second mechanism is due to a current carried by nearly trapped
electrons in a boundary layer close to the trapped-passing boundary
\cite{catto,connor}.  Both of these mechanisms require finite electron
temperature gradient and collisions. However, MTMs have been observed
to arise under conditions which are not well-described by the above
mechanisms \cite{ApplegatePPCF2007,Dickinson}. Thus, a complete
picture of MTM excitation is not available at present.

In the following we summarize the previous findings reported in
Refs. \cite{ApplegatePPCF2007,Dickinson,GuttenfelderPoP2012L,GuttenfelderPoP2012NL,DoerkPoP2012}.
Linear simulations for MAST \cite{ApplegatePPCF2007,Dickinson},
NSTX\cite{GuttenfelderPoP2012L}, and ASDEX-UG \cite{DoerkPoP2012}
plasmas reported that MTMs can be the dominant instabilities in the
region $r/a=0.5-0.8$ with the maximum growth rate at mode numbers
$k_{\theta}\rho_s$ between $0.2-0.8$. Here, $\rho_s$ is the ion sound
Larmor radius.  The electron temperature gradient is found to be the
drive of the instability and a non-monotonic dependence of the growth
rate on the electron-ion collisionality $\nu_{ei}$ is observed. The
peak of the growth rate coincides with the experimental value of
$\nu_{ei}$ for various considered radial positions.  The non-monotonic
dependence of the MTM growth rate on collisionality is due to the fact
that, on one hand both of the driving mechanisms mentioned above
require finite collisionality, and therefore stability can be expected
as $\nu_{ei}$ reduces, but on the other hand, in a strongly
collisional regime the strong rate of scattering of the electrons
between the field lines prevents the formation of a current layer,
hence MTMs are stabilized \cite{DoerkPoP2012}.

It has also been shown that by increasing the effective ion charge
$Z_{eff}$, MTMs are destabilized through the $Z_{eff}$-dependence of
the electron-ion collision frequency, see
Ref.~\cite{GuttenfelderPoP2012L}. As MTMs are electromagnetic in
nature a finite $\beta_e$ is needed for their destabilization. For the
studied discharge in Ref.~\cite{GuttenfelderPoP2012L} it is found that
the $\beta_e$ threshold is well below the experimentally relevant
$\beta_e$ value and the growth rate increases moderately with
increasing $\beta_e$.
 
Non-linear simulations confirmed the role of the electron temperature
gradient as the drive of the MTM instability in spherical (NSTX)
\cite{GuttenfelderPRL2011} as well as standard tokamaks (ASDEX-UG)
\cite{DoerkPoP2012}. Moreover, it has been shown that the Chirikov
criterion \cite{Chirnikov1979PR} for overlapping of the magnetic
islands leads to an up-shift of the electron temperature gradient
threshold
\cite{DoerkPoP2012,DoerkPRL2011,GuttenfelderPoP2012NL,GuttenfelderPRL2011}.

In the present paper we investigate the onset of the MTMs and its
parametric dependence through local linear gyrokinetic simulations
with the \gyro code~\cite{gyro}, in a spherical tokamak: NSTX, and two
conventional tokamaks: ASDEX-UG and JET. For the NSTX case we use the
plasma parameters reported in
Ref.~\cite{GuttenfelderPoP2012L,GuttenfelderPoP2012NL}, and for the
ASDEX-UG case we use the plasma parameters found in
Ref.~\cite{DoerkPoP2012}. In the present work we re-examine these
discharges with an emphasis on the parametric dependences of the MTM
onset. In view of the new ITER-like wall experiments on JET, in the
presented analysis we have also considered a set of JET-like
parameters.

The remainder of the paper is organized as follows. In
Sec.~\ref{sec:input} the input parameters are discussed, and in
Sec.~\ref{sec:1} parametric dependences of the MTM onset are analyzed
by presenting scans over MTM driving parameters such as
collisionality, $\beta_e$ and electron temperature/density scaling
lengths. The conclusions are drawn in Sec.~\ref{sec:conclusions}.

\section{Input parameters}
\label{sec:input}
The plasma parameters used in our analysis are shown in table \ref{table1}. 
\begin{table}[ht]
\caption{Input parameters for densities, temperatures and their gradients.}  
\centering
\begin{tabular}{|c|c|c|c|c|c|c|c|c|c|c|c|c|c|c|}
\hline\hline
            & $r/a$& $Z_{eff}$ &  $n_e [10^{19}/m^3]$ &$T_e[keV]$ &$a/L_{ne}$ & $a/L_{Ti}$ & $a/L_{Te}$ &$T_i/T_e$ & $\rho_s/a$& $\nu_{ei} (a/c_s)$\\ [0.5ex] \hline
NSTX & 0.6     & 	     2.91        & 	6.0 		     & 0.44            & -0.83           & 2.36            & 2.72            & 0.94           & 0.0074       &        1.45  		    \\ 
\hline
AUG   & 0.65   &             3.30        &     7.6		     &  0.765         & 0.37            & 2.18            & 3.02            & 1.19           & 0.0018       &        0.68		    \\
\hline         
JET     & 0.6     &              3.41       & 	7.8 	    	     & 1.25            & 0.15            & 2.16            & 2.16            & 1.00           & 0.0027       &        0.43		    \\
\hline
\end{tabular}
\label{table1}
\end{table} 

Here, $L_{n}=-[\partial (\ln{n})/\partial r]^{-1}$, $L_{T}=-[\partial
(\ln{T})/\partial r]^{-1}$, are the density and temperature scale
lengths, $a$ is the outermost minor radius.
\begin{table}[ht]
\caption{Input parameters for plasma ion compositions. }  
\centering
\begin{tabular}{|c|c|c|c|c|c|c|c|c|c|c|c|c|c|c|}
\hline\hline
            & $Z_1$      &  {$n_{Z1}/n_e$}       &${a/L_{nZ1}}$ & $Z_2$          & ${n_{Z2}/n_e}$\\ [0.5ex] \hline
NSTX & $C^{+6}$ &   6.4$\%$                     & -2.75                       & $W^{+40}$ &0.02$\%$           \\ 
\hline
AUG   & $N^{+7}$ &   4.8$\%$                     & 0.80                        & $W^{+40}$  &0.02$\%$           \\
\hline
JET     & $N^{+7}$ &   5.0$\%$                     & 0.14                        & $W^{+40}$  &0.02$\%$           \\
\hline
\end{tabular}
\label{table2}
\end{table}
Deuterium ions, an active impurity species denoted in table
\ref{table2} by $Z_1$ (carbon for NSTX and nitrogen for ASDEX-UG and
JET), and a passive species of impurity (tungsten unless otherwise
stated) denoted in the table \ref{table2} by $Z_2$, are considered. The
passive species are considered here in order to examine the impurity
particle transport due to MTMs.  Note that in the ASDEX-UG case
reported in Ref.~\cite{DoerkPoP2012} no impurities were present and
the value of $a/L_{Ti}$ has been artificially reduced from its
experimental value to eliminate the drive of Ion Temperature Gradient
(ITG) modes, but here we use the experimental values.

Linear runs with \gyro include full electromagnetic effects: shear
$\delta B\;(=\nabla\times\delta A_{\parallel})$, and compressional $\delta
B_{\parallel}$ magnetic perturbations. Gyrokinetic electrons are
assumed, and the collisions are modeled using an energy dependent
Lorentz operator.  Both electron-ion and electron-electron collisions
are included in the electron collision frequency $\nu_e(v)$, and
collisions between all ion species are accounted for. To take into
account the plasma shape we have used a Miller-type local equilibrium
model available in {\sc gyro}, see Refs.~\cite{waltz09,candy09}. 
Typical JET parameters for plasma shape and magnetic geometry are
used, and the corresponding values are given in Table~\ref{table3}.
\begin{table}[ht]
\caption{Input parameters for plasma shape and magnetic geometry.}
\centering
\begin{tabular}{|c|c|c|c|c|c|c|c|c|c|c|c|}
\hline\hline &$a [m]$& $\beta_e$ &$\alpha_{MHD}$ &$B_{unit}$& $R_0/a$ & $B_{0} [T]$ & $s$   & $q$  & $\kappa$ & $\delta$ & $k_{\theta}\rho_{s}$\\ [0.5ex] \hline 
NSTX          &0.6        & 0.024         & 0.36			 & 0.66          & 1.52               &  0.35                   & 1.73 & 1.68 & 1.72          & 0.12        & 0.63               \\
\hline 
AUG            &0.6        & 0.005         & 0.42		 	 & 2.16          & 3.3                 &  2.479                 & 1.31 & 2.18 & 1.30          & 0.13        & 0.2                  \\ 
\hline
JET              &1.0        & 0.013         & 0.35             	  &1.77           & 3.3                 & 2.55                    & 1.32 & 1.45 & 1.70          & 0.37        & 0.5                  \\ 
\hline
\end{tabular}
\label{table3}
\end{table}
In this table, $\beta_e$ is calculated in CGS units following the
expression:
\begin{equation}
\beta_{e}=\frac{8 \pi ( n_e[10^{19}/m^3]\;10^{-6} \;10^{19} )(
 T_e[keV]\;1.6022\;10^{-9} )}{(10^{4}\; B_{unit}[T] )^2},
\label{betae}
\end{equation}
where $B_{unit}=(d \chi_t/dr)/r$ is the effective field strength
with $\chi_t$ defined through the toroidal magnetic flux $2 \pi
\chi_t$ \cite{waltz09,candy09}, $q$ is the safety factor, and the
magnetic shear is $s=(r/q)dq/dr$. The generalized magnetohydrodynamic $\alpha$ parameter is defined as
\begin{equation} 
\alpha_{MHD}=-q^2 R_{0}\frac{8\pi}{B_{unit}^2}\frac{dp}{dr}c_p
,\label{alpha}
\end{equation}
where $R_{0}$ is the effective major radius, $r$ is the minor radius,
and $p=\sum_{a}n_{a}T_{a}$ is the total plasma pressure. $c_{p}$ is a
scaling parameter which allows an artificial adjustment of
$\alpha_{MHD}$ (affecting the magnetic curvature drift) without
modifying the background gradients, as presented in Ref.~\cite{belli}.
Furthermore, $\kappa$ is the elongation, $\delta$ is the
triangularity. 

Typical resolution parameters used in our linear analysis are as
follows: 40 radial grid points, 12 parallel orbit mesh points ($\times
2$ signs of parallel velocity), 16 pitch angles, and 8 energies. A
high radial resolution is needed since the linear instability of MTM
depends on the presence of narrow resonant current layers centered on
the rational surfaces. The toroidal mode numbers used here,
corresponding to the $\rho_s/a$ values in table \ref{table1}, are
$n=30$ for NSTX, $n=18$ for AUG, and $n=75$ for JET.

\section{MTM instability}
\label{sec:1}
\begin{figure}[htbp]
\begin{center}
  \includegraphics[width=1.\textwidth]{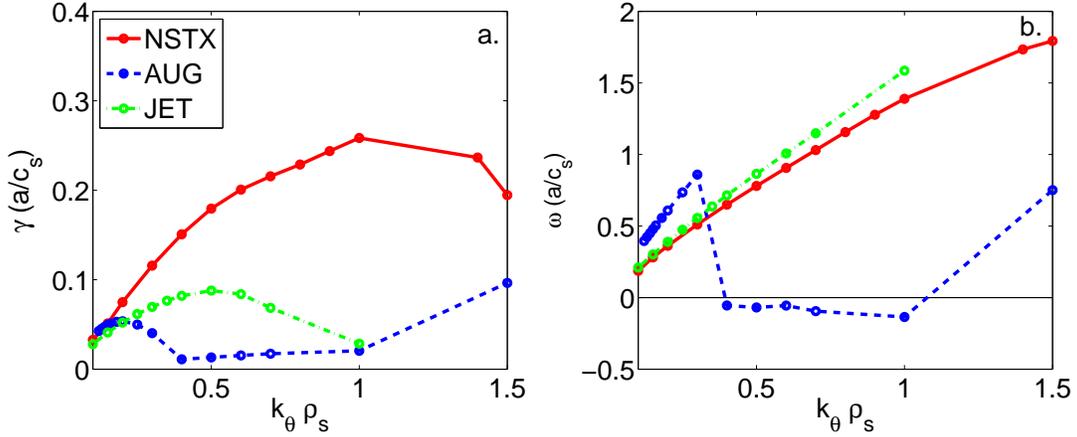}
  \caption{Imaginary (a), and real parts (b) of eigenvalues ($\gamma$,
    $\omega_{r}$) as functions of $k_{\theta}\rho_{s}$. Red solid
    lines: NSTX, blue dashed: ASDEX-UG, green dash-dotted:
    JET.}
\label{eigenvalues}
\end{center}
\end{figure}

Figures~\ref{eigenvalues} (a,b) illustrate the linear growth rate and
the real frequency of the most unstable modes in the three machines as
functions of $k_{\theta}\rho_{s}$. In NSTX and JET, the MTMs (with
positive real frequency in electron diamagnetic direction) are the
most unstable modes in the range of $0.1 \le k_{\theta}\rho_{s} \le
1$, while for ASDEX-UG the MTMs are the most unstable modes only in
the narrower wave number range $0.1 \le k_{\theta}\rho_{s} \le 0.3$,
corresponding to longer wavelengths than those in NSTX and
JET. 
{In NSTX the MTMs remain the most unstable mode
  for the whole considered range of $k_{\theta}\rho_{s}$, but in
  ASDEX-UG an ITG mode (negative real frequency, in ion diamagnetic
  direction) for $0.3<k_{\theta}\rho_{s}<1$ and an ETG mode (positive
  real frequency, in electron diamagnetic direction) for
  $1<k_{\theta}\rho_{s}$ are also found to be unstable.} The
$k_{\theta}\rho_{s}$ corresponding to the maximum growth rate varies
between the different machines: in NSTX $k_{\theta}\rho_{s}\sim0.6$,
in ASDEX-UG $k_{\theta}\rho_{s}\sim 0.2$ and in the JET like case
$k_{\theta}\rho_{s}\sim 0.5$. The normalized poloidal mode number for
the remainder of our calculations are fixed to these values (see table
\ref{table3}). Previous studies have discussed the difference in the
poloidal mode number corresponding to the maximum of the unstable MTMs
between the NSTX and ASDEX-UG cases, and it is believed to be due to
characteristics of the spherical or conventional tokamaks
\cite{DoerkPoP2012,GuttenfelderPoP2012L}. However, here we find
unstable MTMs with similar mode numbers in the JET tokamak to that of
NSTX, which suggests that this is not always the case. Therefore,
further studies are needed to determine the reason for the
similarities and differences in the mode numbers for maximum growth
rates in various machines.

\begin{figure}[htbp]
\begin{center}
 \includegraphics[width=1\textwidth]{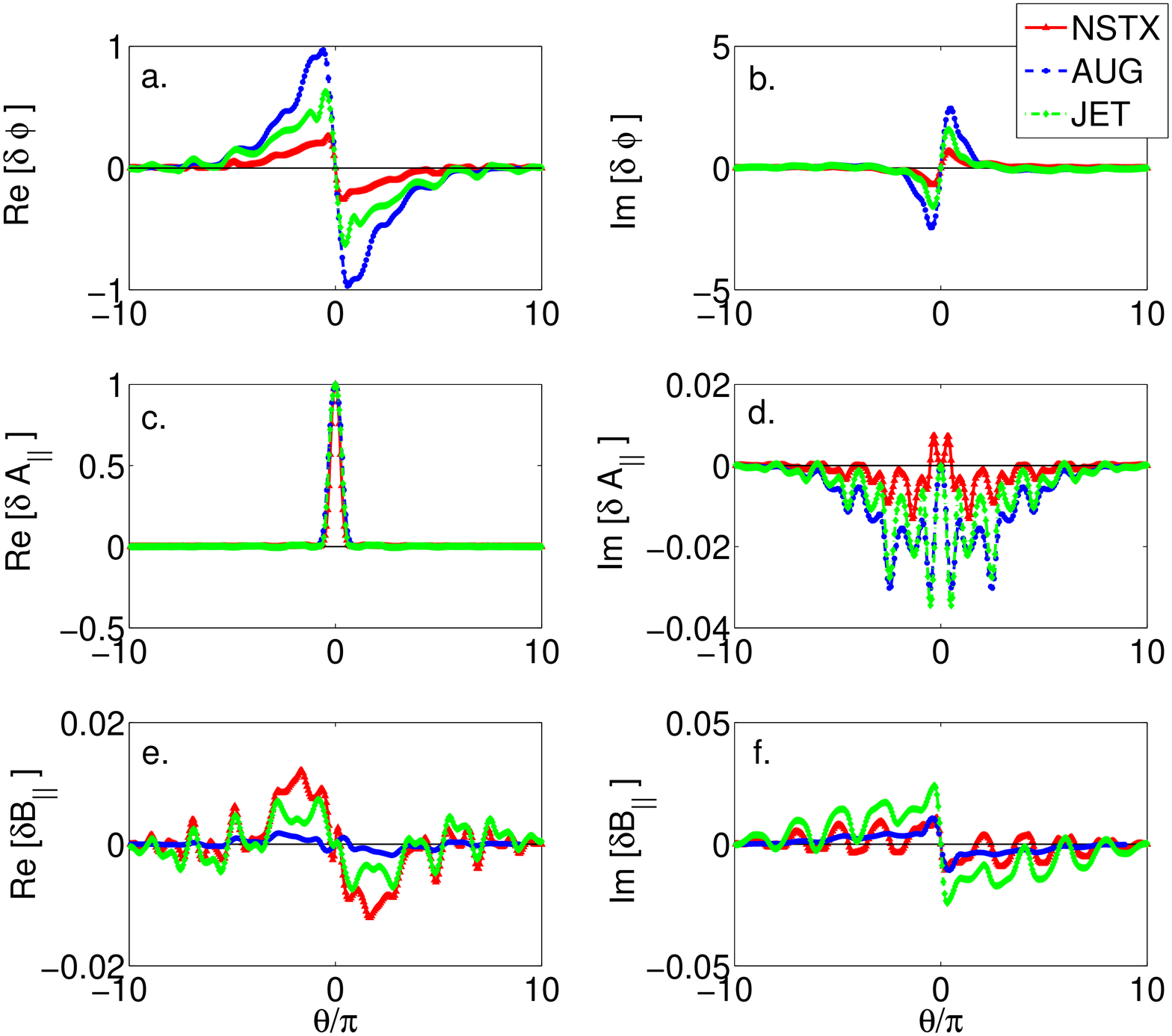}
  \caption{ Linear parallel mode structures of $\delta \phi$ (a,b),
    $\delta A_{\parallel}$ (c,d) and $\delta B_{\parallel}$ (e,f).
    Note that the actual radial resolution of the simulations covers
    $\theta/\pi=(-16,16)$.}
\label{modestructure}
\end{center}
\end{figure}

The structure of the $\delta \phi$, $\delta A_{\parallel}$ and $\delta
B_{\parallel}$ eigenmodes, corresponding to the $k_{\theta}\rho_s$
values mentioned above are shown in figures \ref{modestructure}
(a-f). The eigenfunctions are normalized so that $\delta
A_{\parallel}(\theta=0)$ is unity. The MTM signature is distinguished
by the tearing parity of the $\delta A_{\parallel}$ eigenmodes.  The
NSTX case has the strongest electromagnetic character in terms of the
relative amplitude of $\delta \phi$, while the ``most electrostatic''
mode is the AUG case. Regarding the strength of the compressional
magnetic perturbations, $\delta B_{\parallel}$, the situation is the
opposite; the less electrostatic the mode is the stronger the $\delta
B_{\parallel}$ perturbations are.  The characteristic width of the
eigenmodes along the field line is similar in all three machines.

\begin{figure}[htbp]
\begin{center}
  \includegraphics[width=0.33\textwidth]{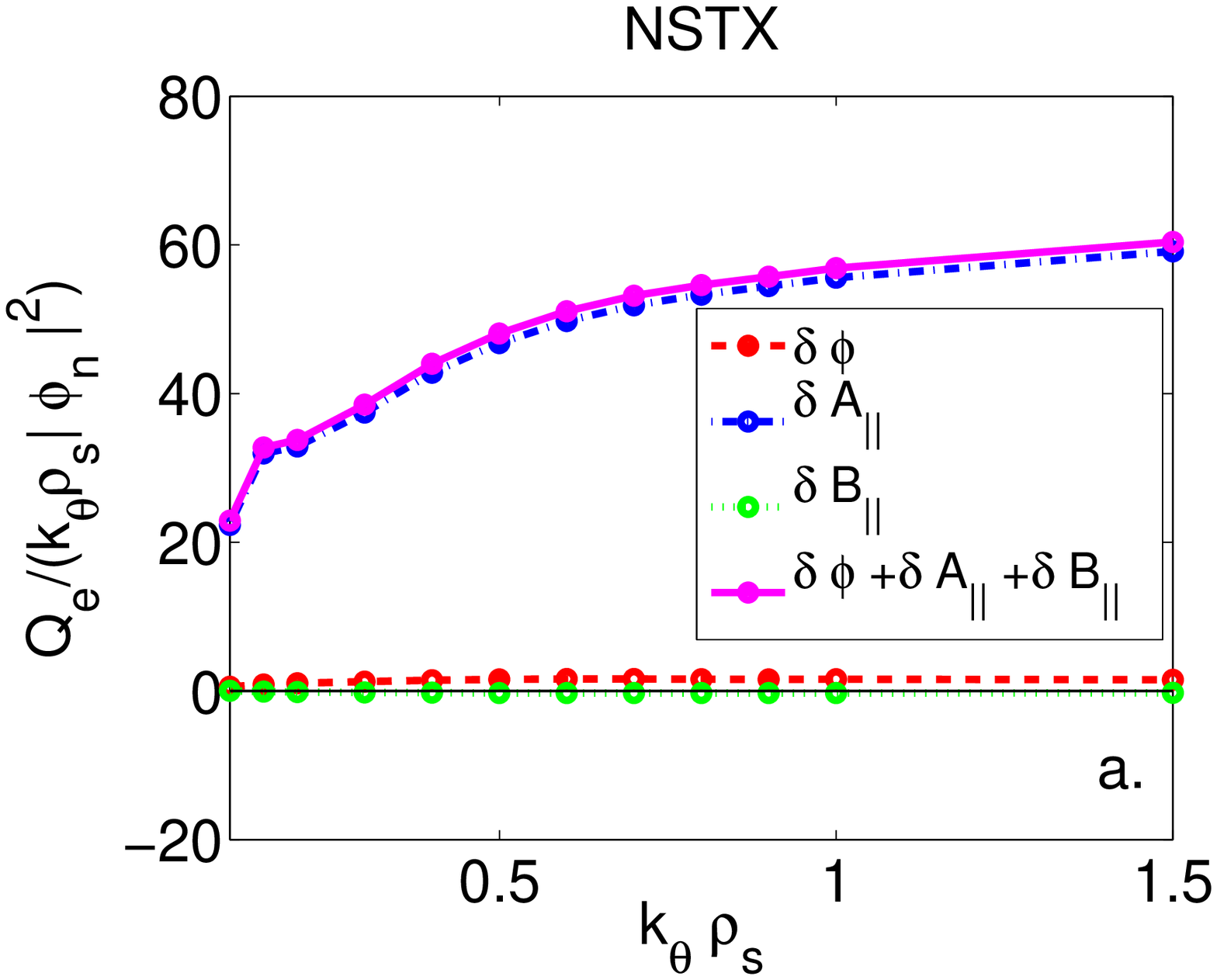} \includegraphics[width=0.33\textwidth]{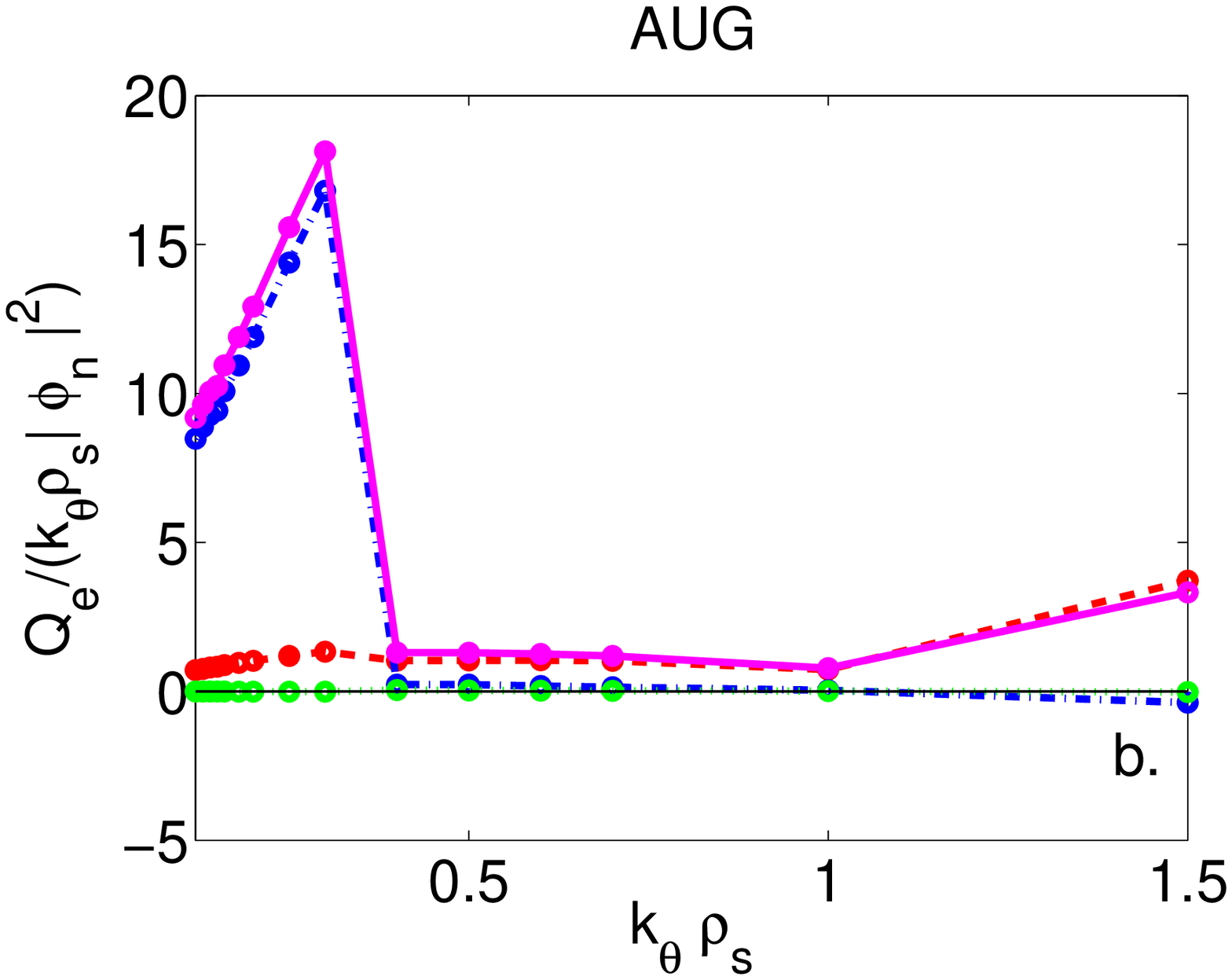}\includegraphics[width=0.33\textwidth]{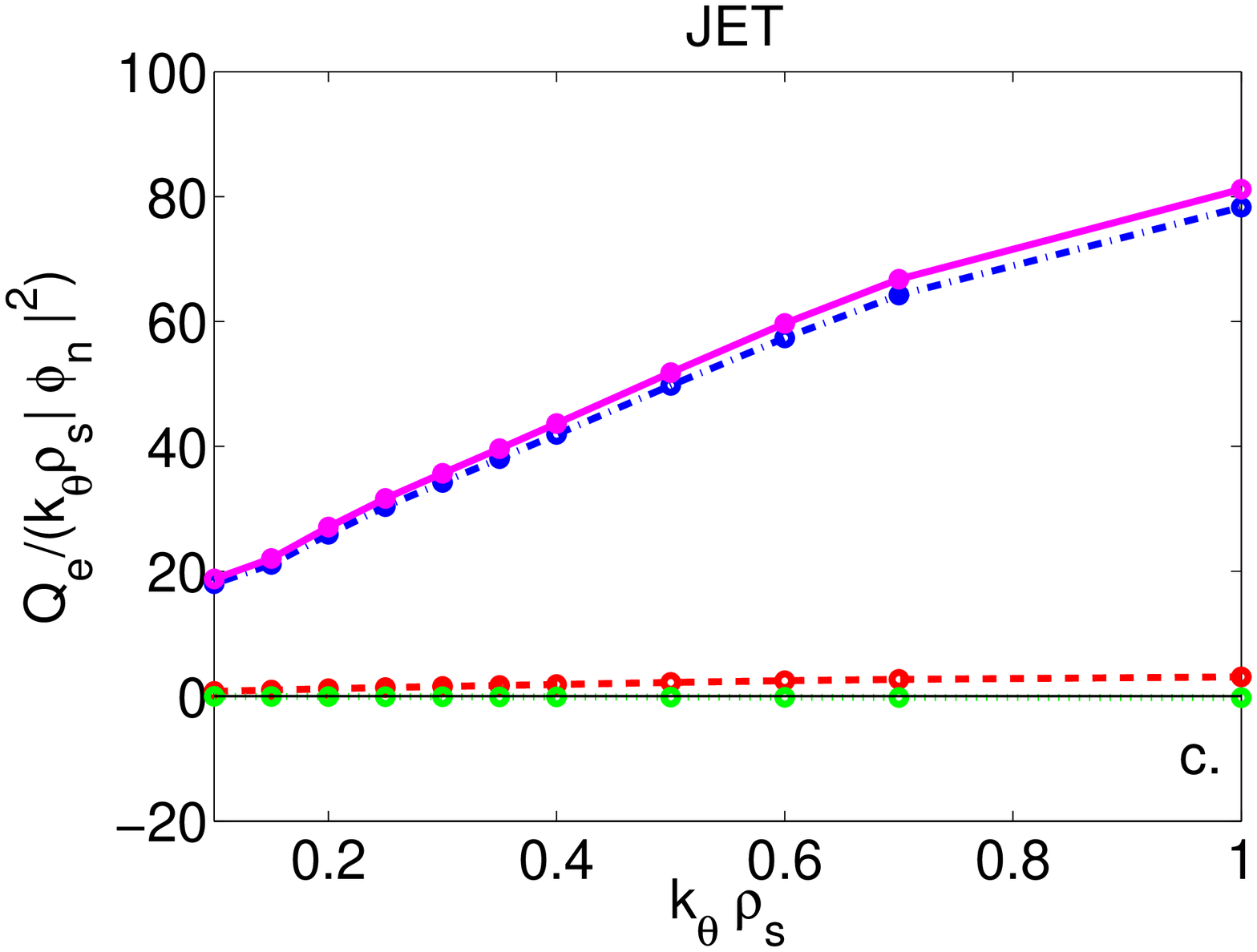}\\
   \includegraphics[width=0.33\textwidth]{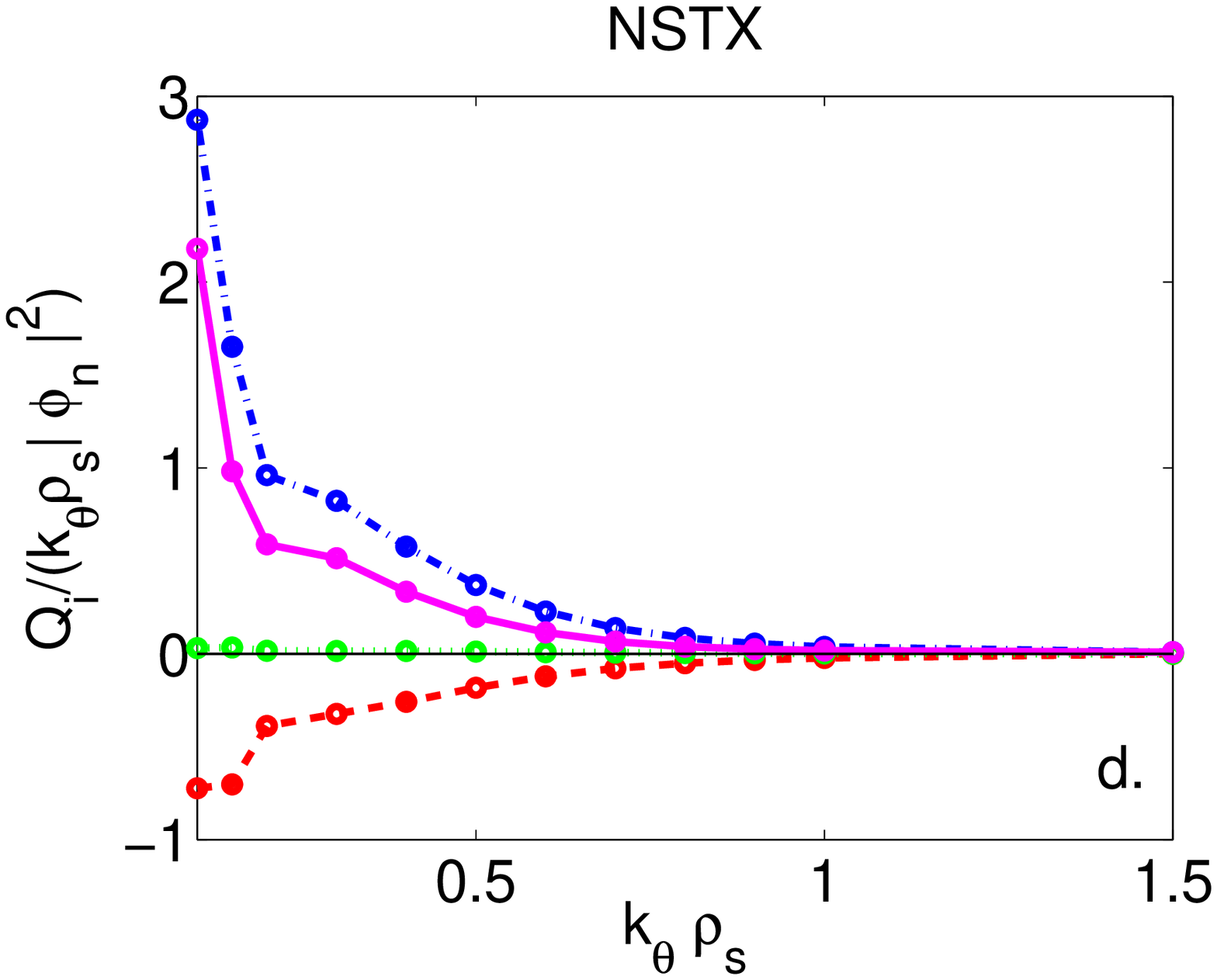} \includegraphics[width=0.33\textwidth]{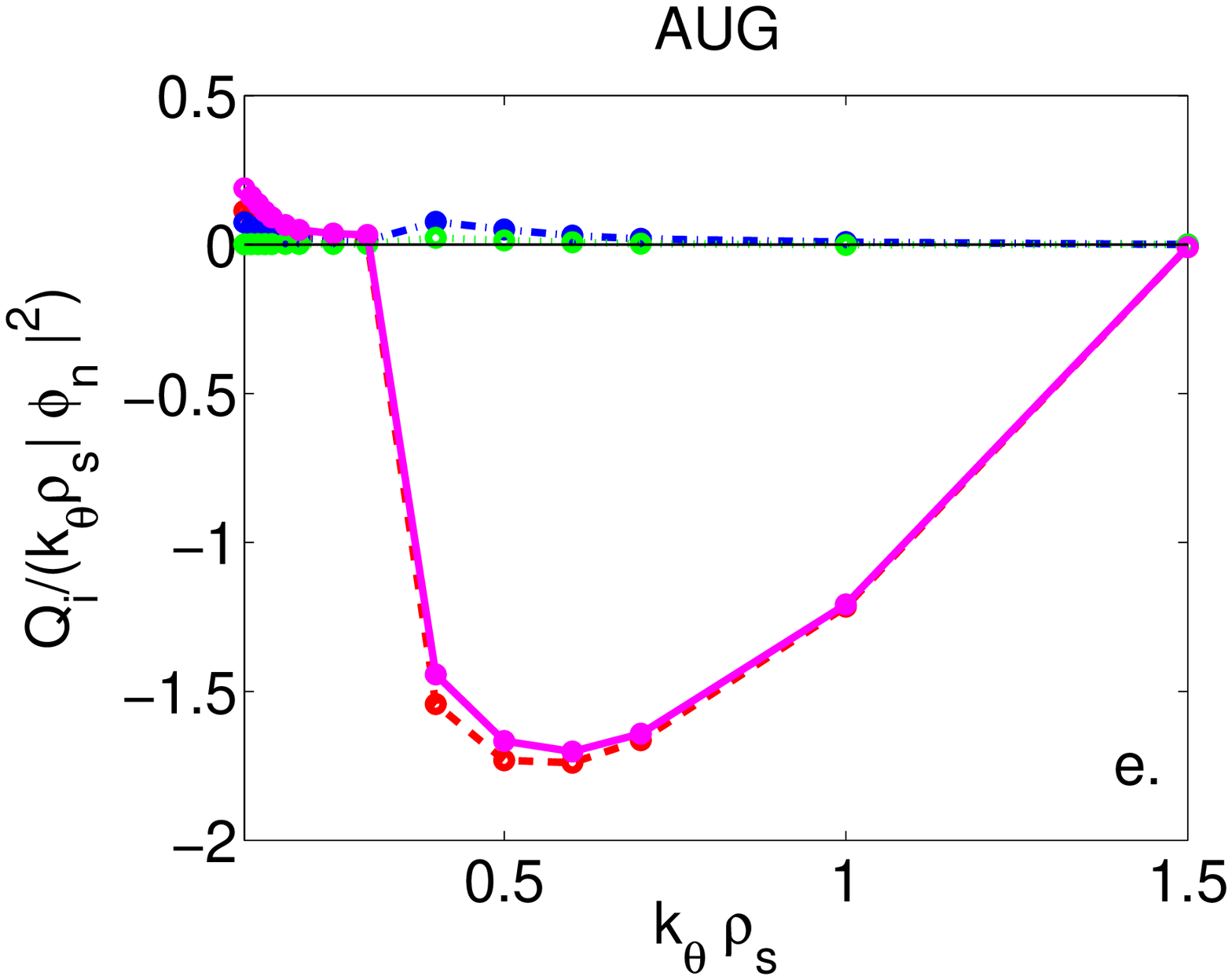}\includegraphics[width=0.33\textwidth]{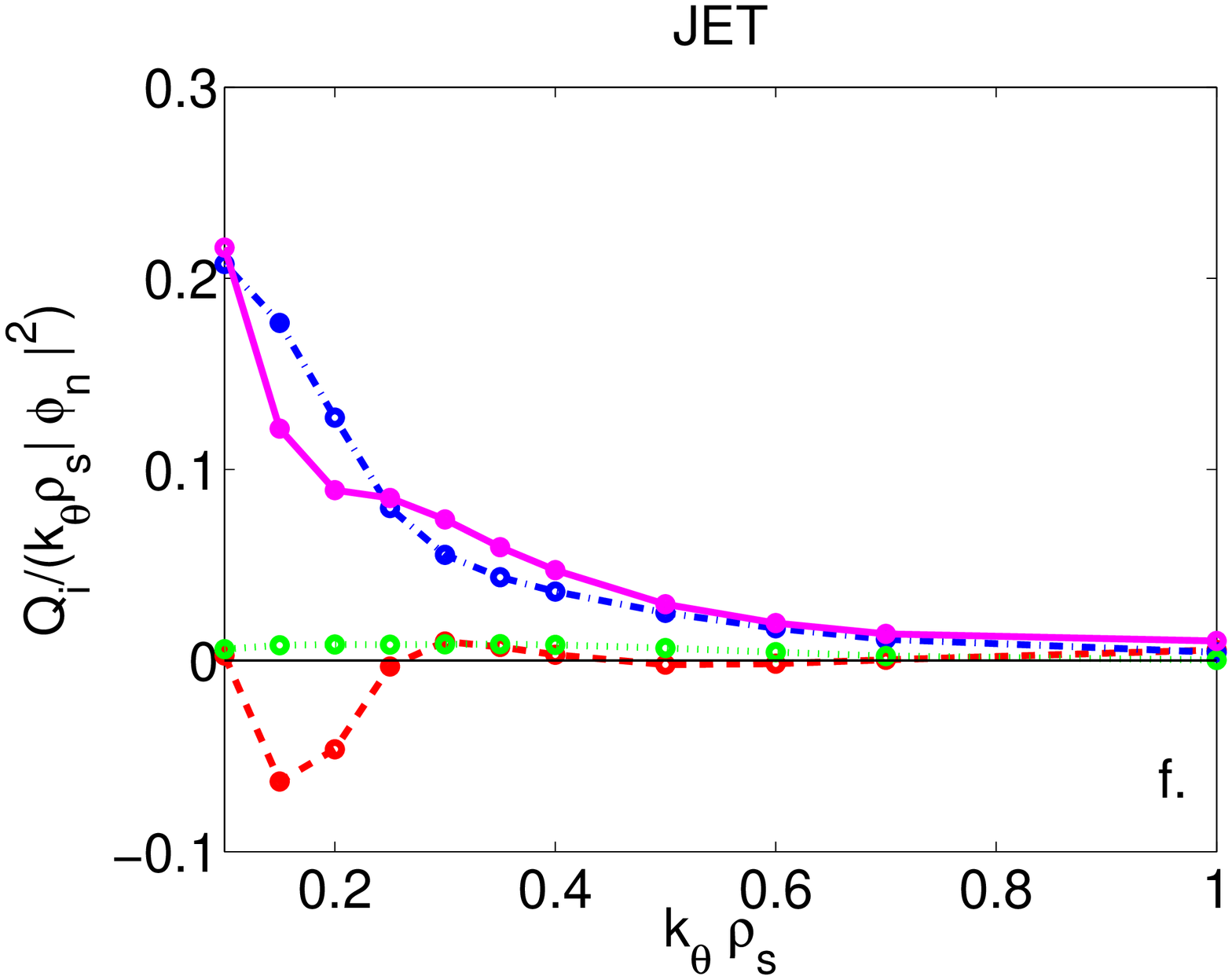}
   \caption{Normalized linear electron (top) and ion (bottom) heat fluxes (magenta solid lines) and
     their contributions from $\delta \phi$ (red dashed lines),
     $\delta A_{\parallel}$ (blue dash-dotted lines), and $\delta
     B_{\parallel}$ (green dotted lines) versus $k_{\theta}\rho_{s}$
     in NSTX (a, d), ASDEX-UG (b, e), and JET
     (c, f). }
\label{heatfluxes}
\end{center}
\end{figure}

\begin{figure}[htbp]
\begin{center}
    \includegraphics[width=0.33\textwidth]{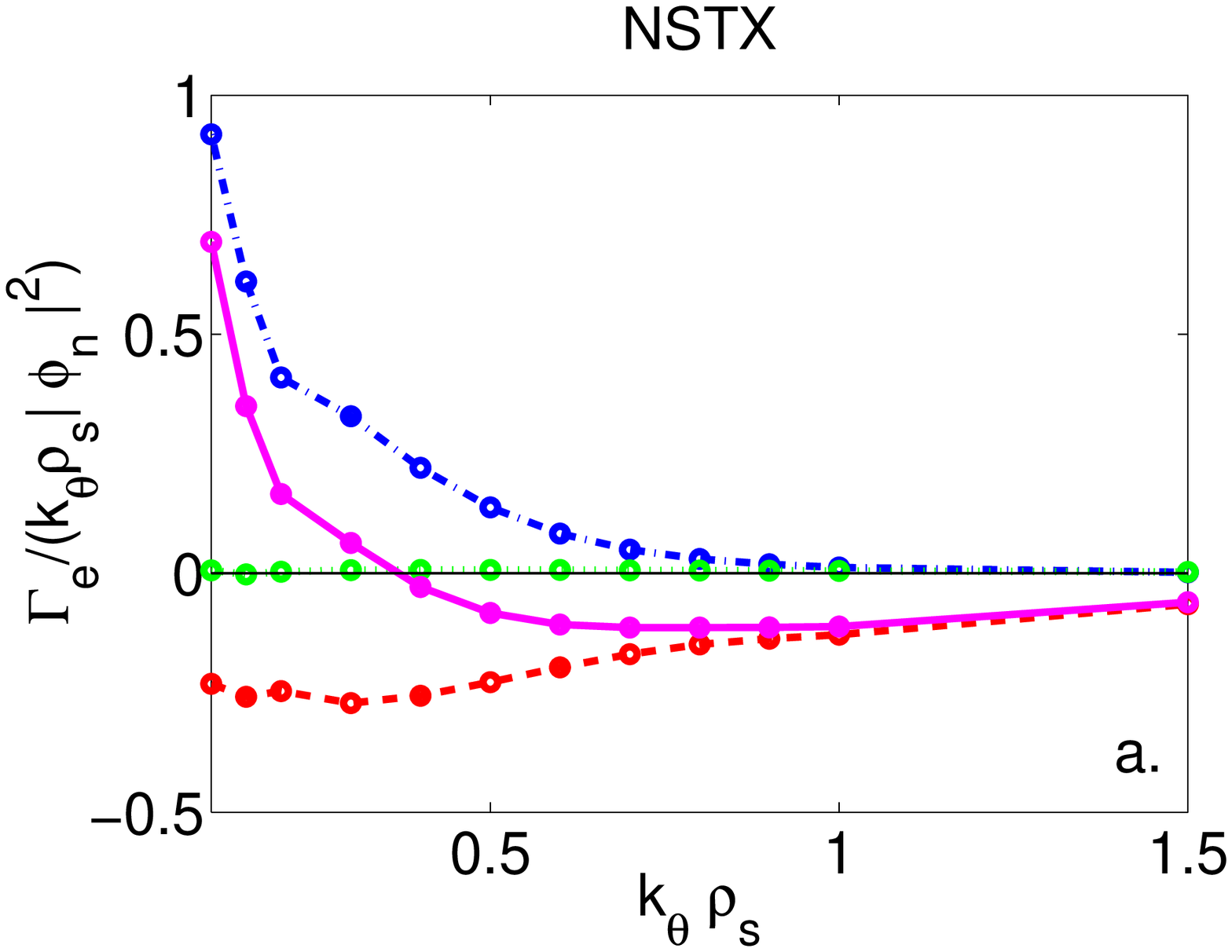} \includegraphics[width=0.33\textwidth]{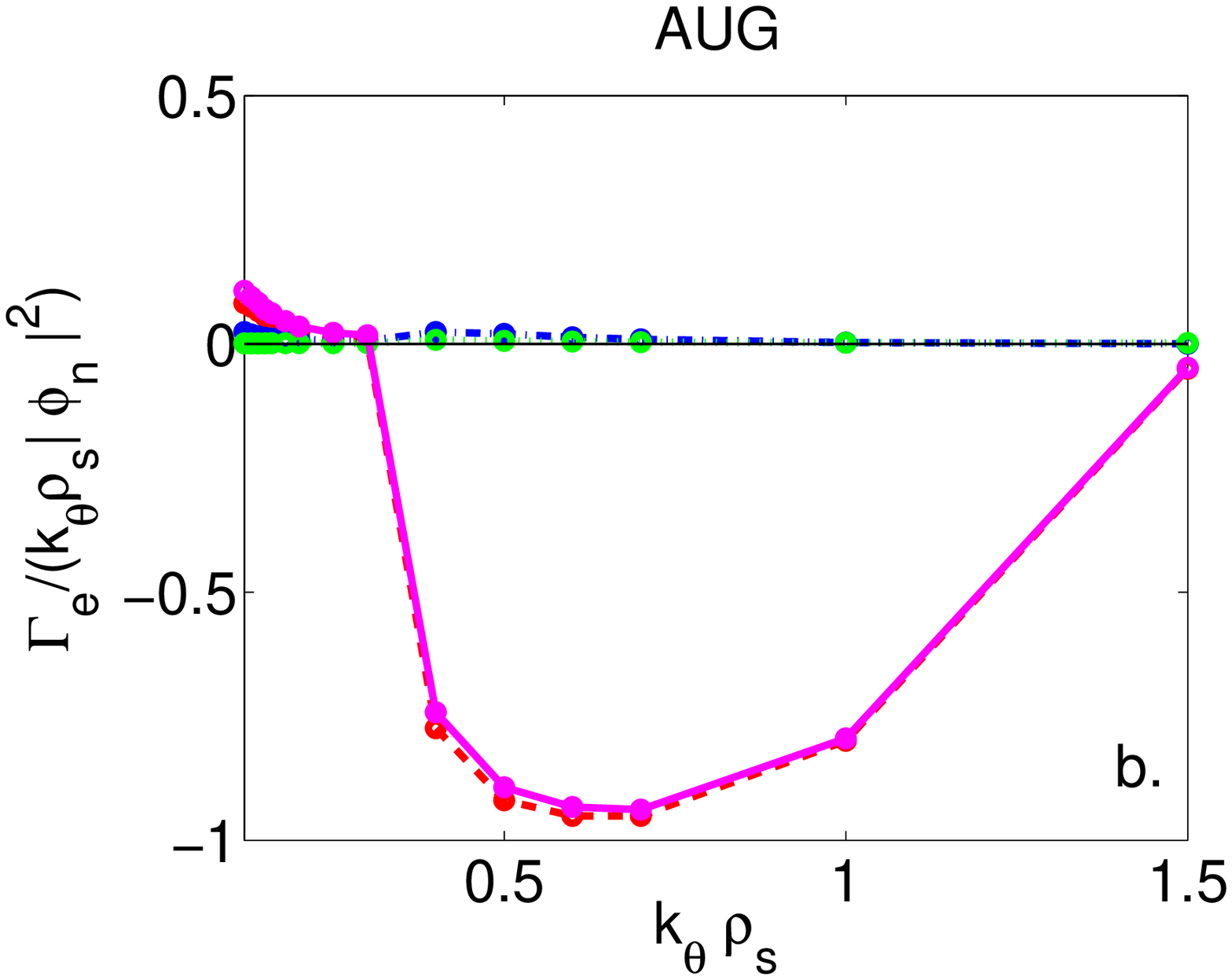}\includegraphics[width=0.33\textwidth]{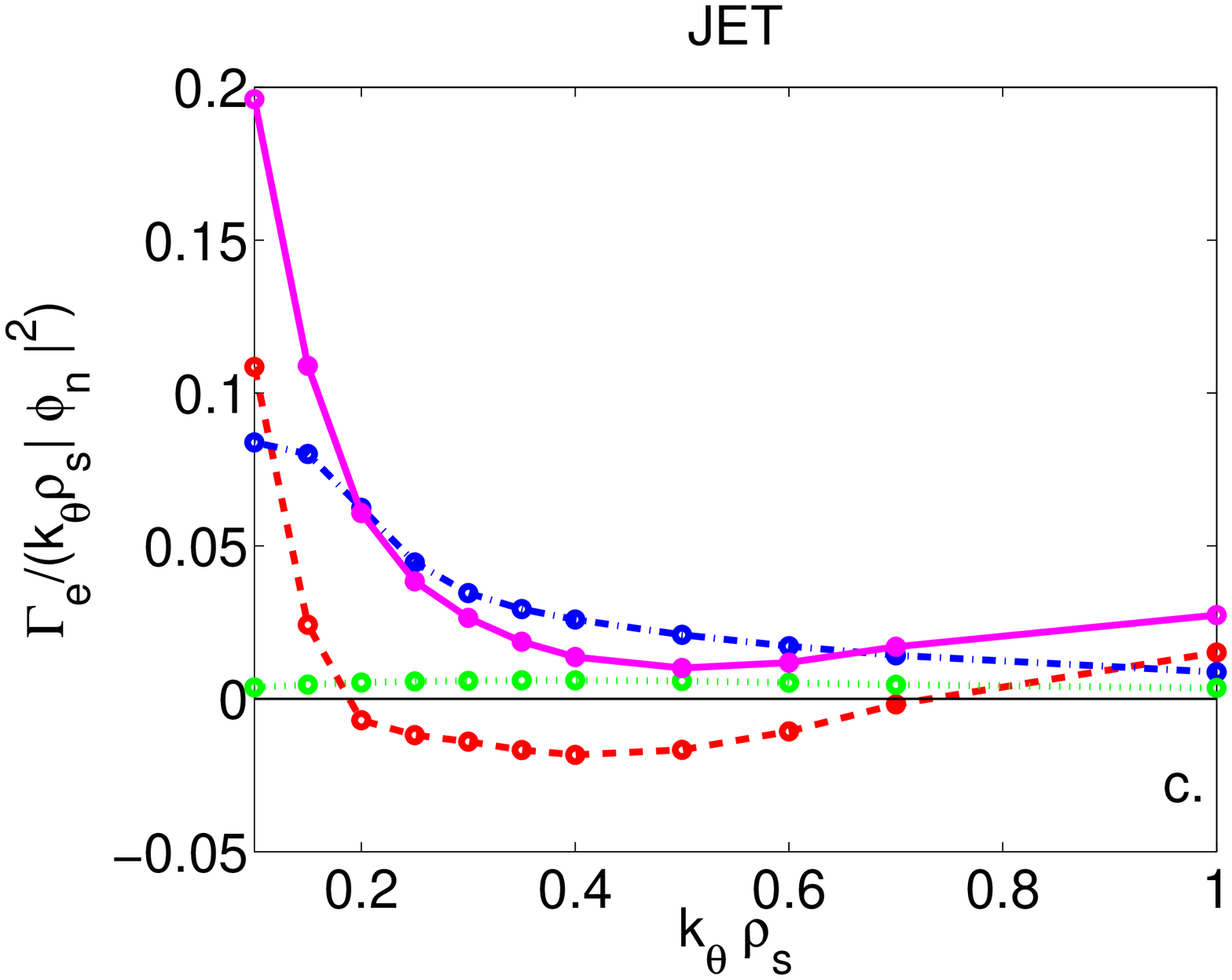}\\
        \includegraphics[width=0.33\textwidth]{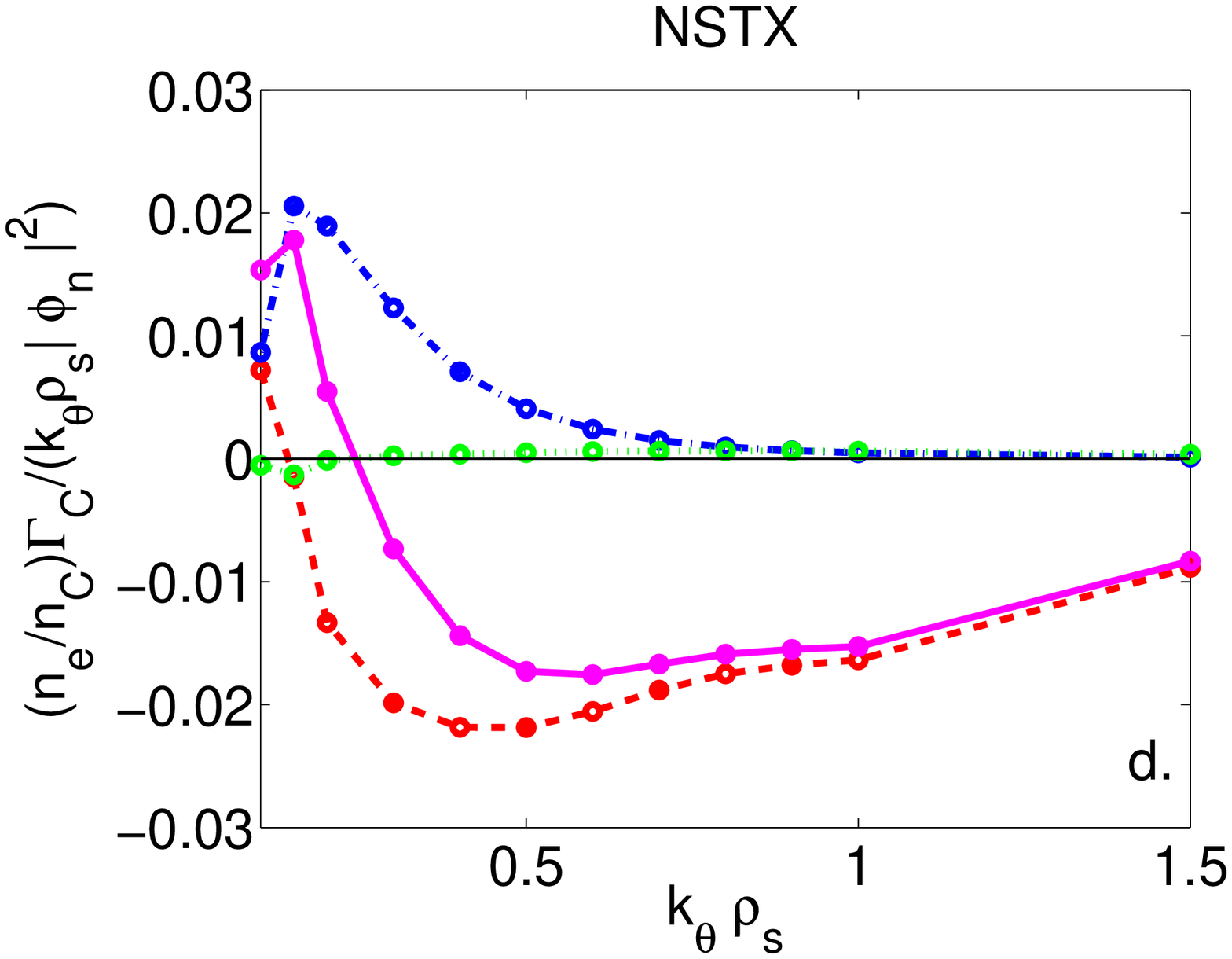} \includegraphics[width=0.33\textwidth]{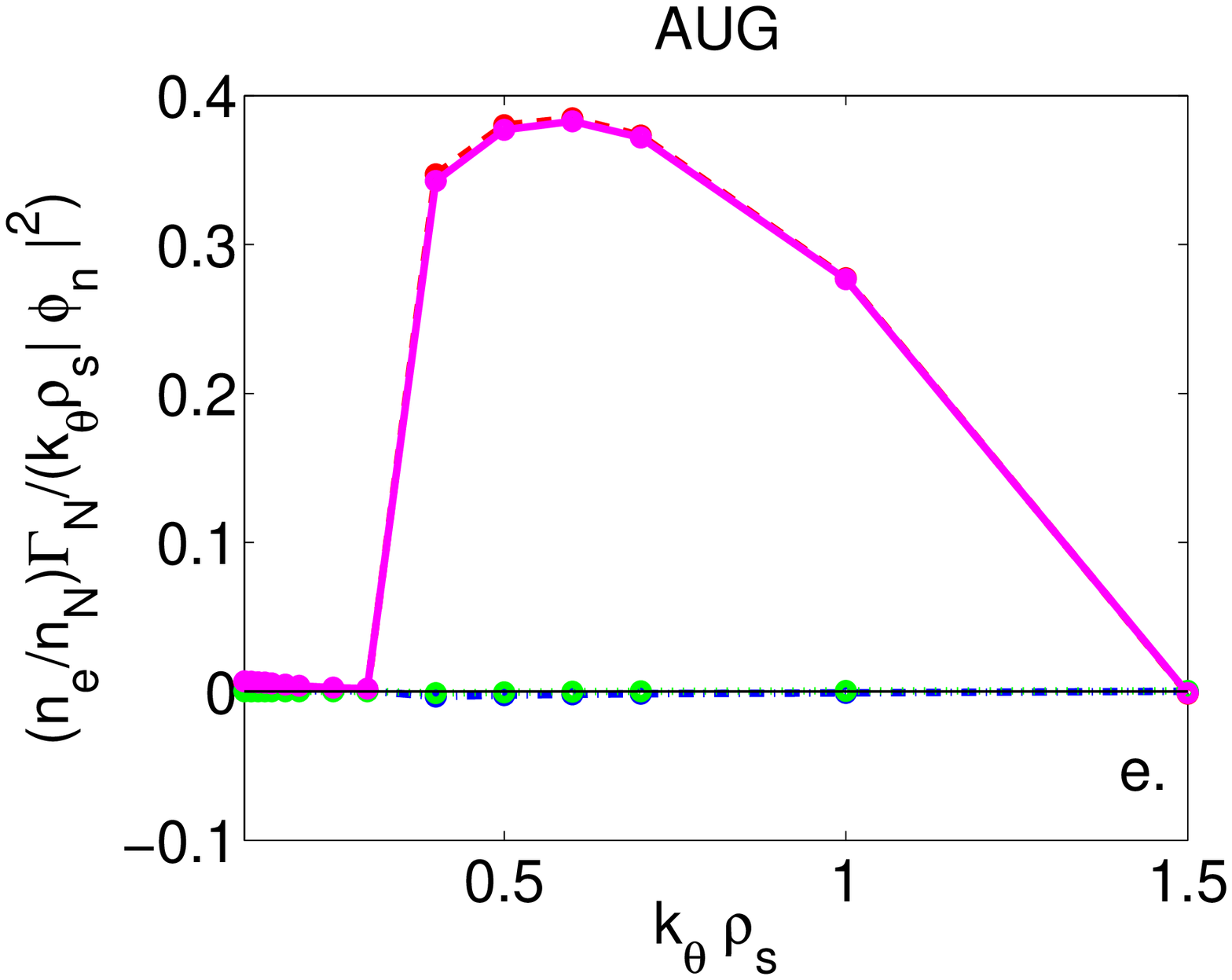}\includegraphics[width=0.33\textwidth]{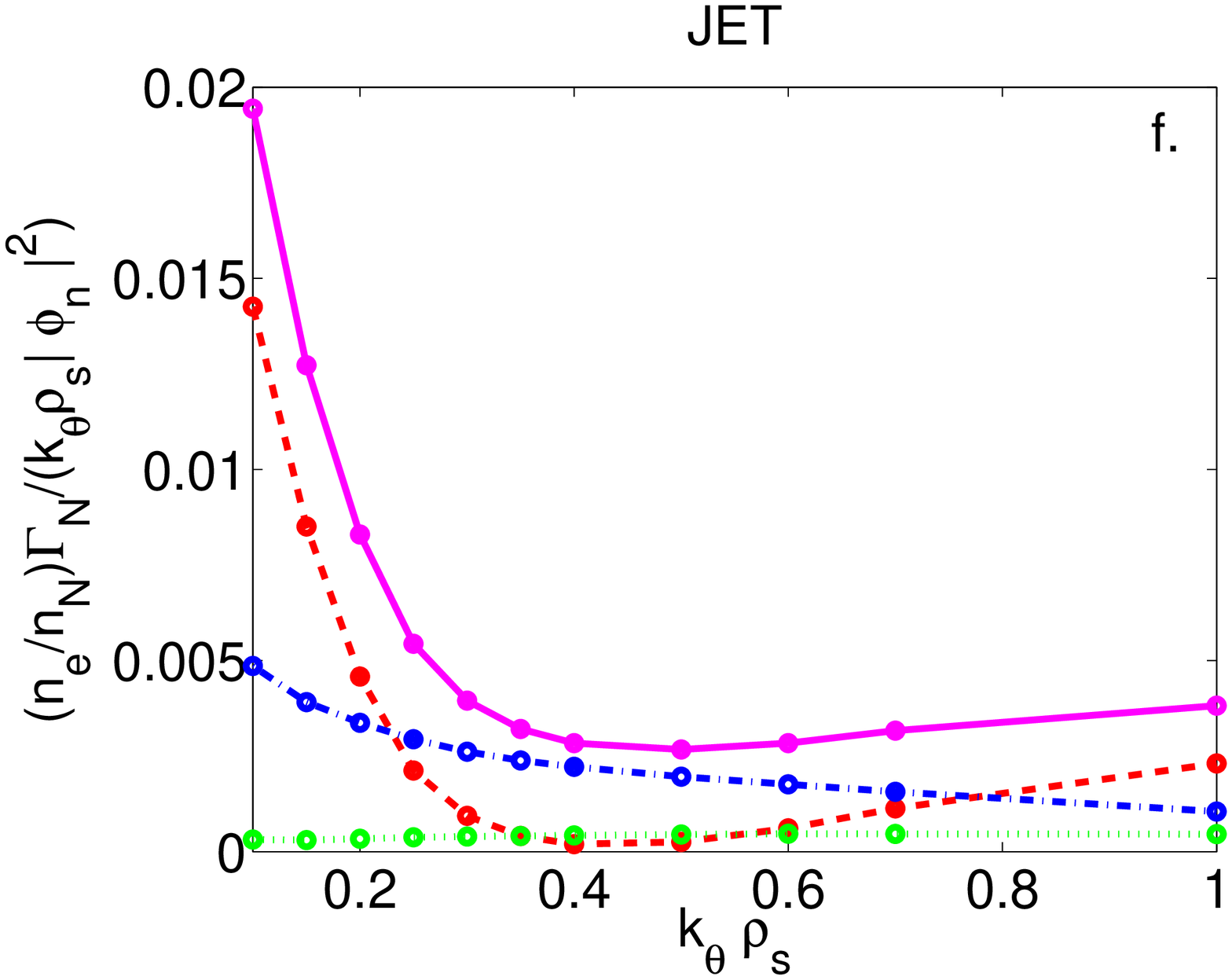}\\
            \includegraphics[width=0.33\textwidth]{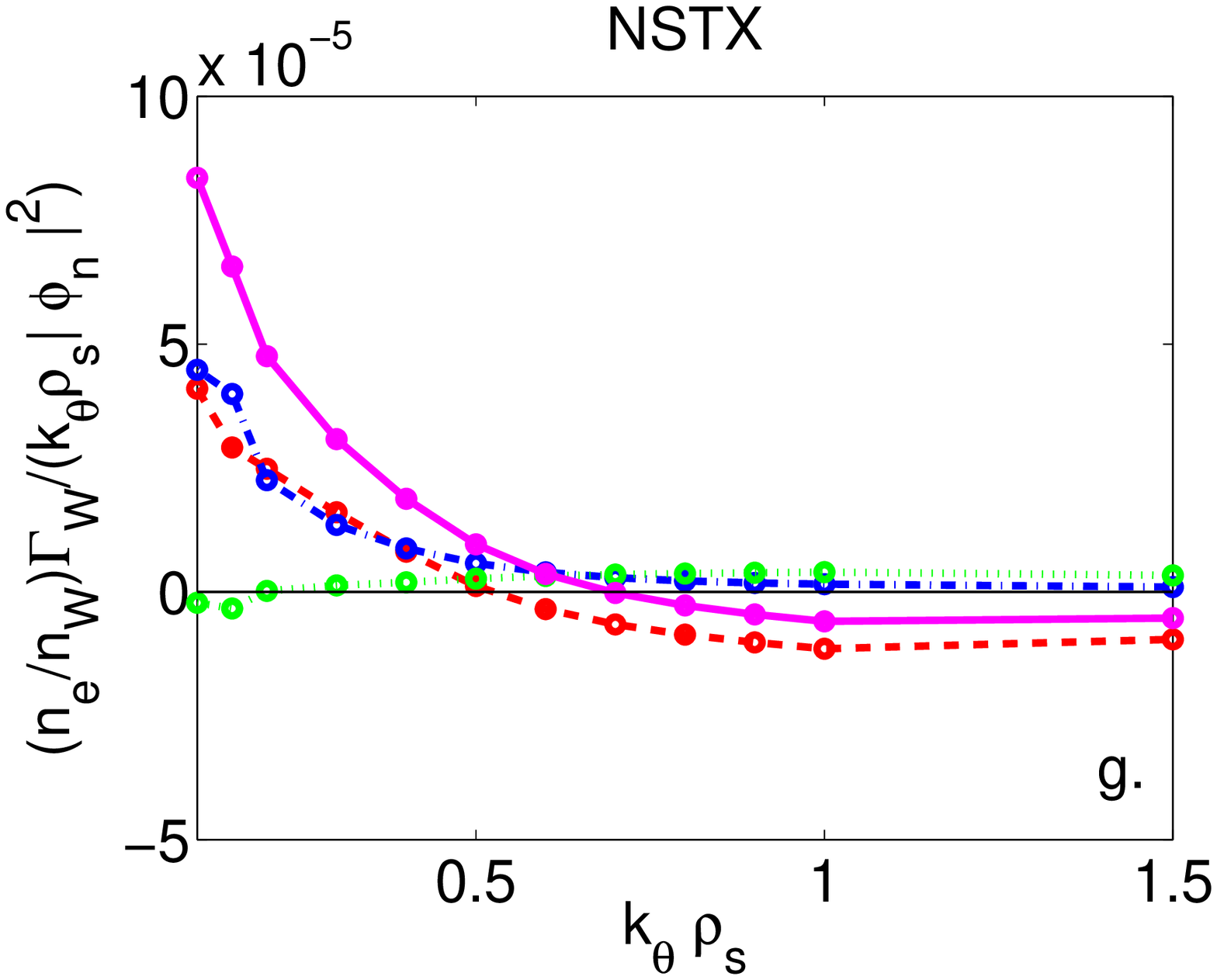} \includegraphics[width=0.33\textwidth]{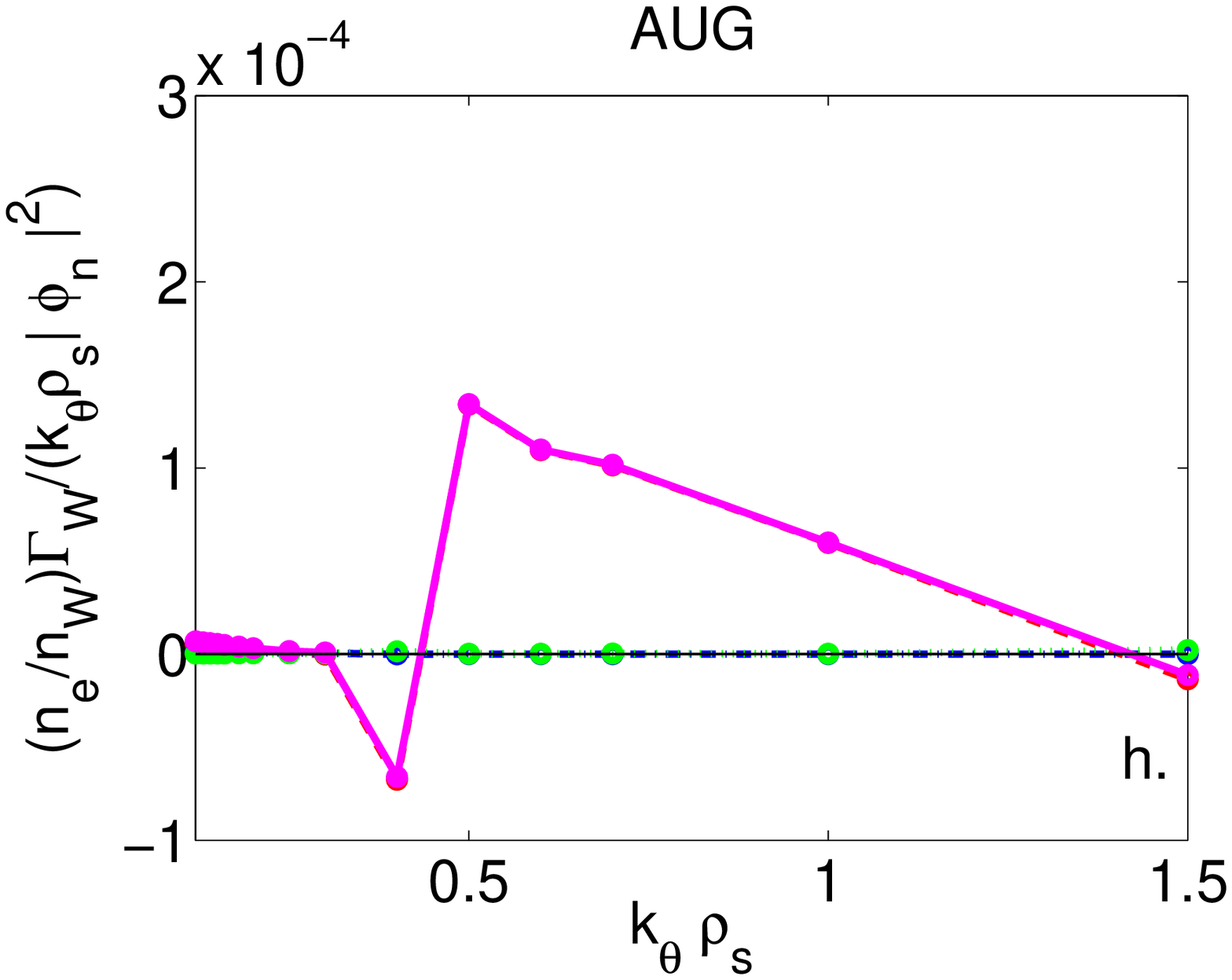}\includegraphics[width=0.33\textwidth]{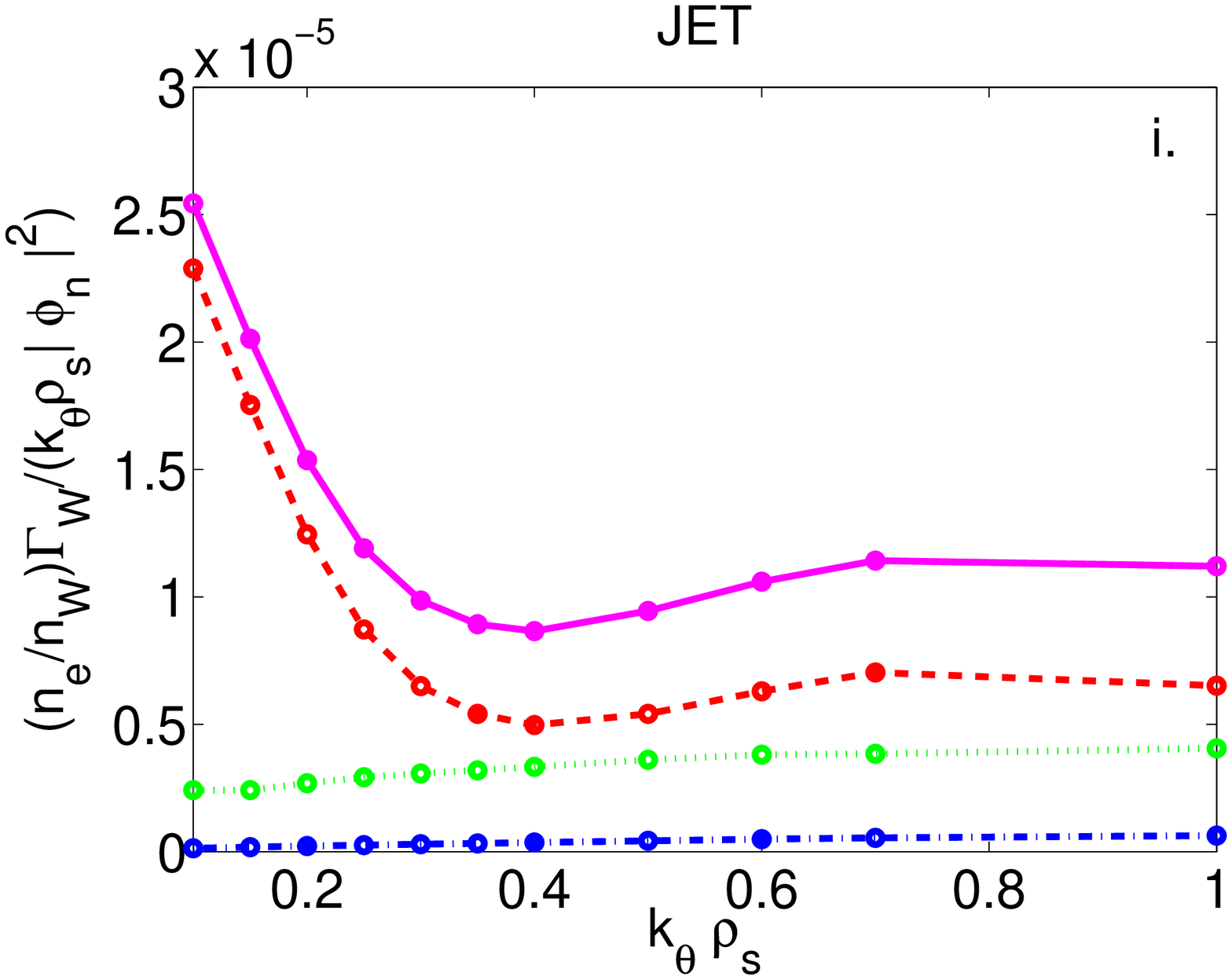}
            \caption{Normalized linear particle fluxes (magenta solid
              lines) and their contributions from $\delta \phi$ (red
              dashed lines), $\delta A_{\parallel}$ (blue
              dash-dotted lines), and $\delta B_{\parallel}$ (green
              dotted lines) versus $k_{\theta}\rho_{s}$ in NSTX (a, d
              and g), ASDEX-UG (b, e and h), and JET (c, f and i).}
\label{particlefluxes}
\end{center}
\end{figure}

Figures \ref{heatfluxes} and \ref{particlefluxes} illustrate the
normalized linear energy and particle fluxes, respectively, and the
contributions to these from the $\delta \phi$, $\delta A_{\parallel}$
and $\delta B_{\parallel}$ fluctuations, as functions of
$k_{\theta}\rho_{s}$ in all three machines. As seen in figures
\ref{heatfluxes} (a-c), in the MTM dominated region the electron heat
flux is the main channel of the energy transport, and the dominant
contribution is generated by $\delta A_{\parallel}$ (blue dash-dotted
line). In the NSTX case, see figure \ref{heatfluxes} (a), this is true
for the whole considered range of $k_{\theta}\rho_{s}$. The ion heat
flux generated by MTMs shown in figure \ref{heatfluxes} (d) (magenta
solid line) however, is negligibly small compared to the electron heat
flux. The particle fluxes generated by MTMs for electrons as seen in
figure \ref{particlefluxes} (a-c), and both active and passive
impurity species shown in figures \ref{particlefluxes} (d-f and g-i)
are also negligible in comparison to the electron heat flux. We note,
that nonlinear MTM simulations for NSTX, presented in
\cite{GuttenfelderPRL2011,GuttenfelderPoP2012NL} also showed
negligibly small particle fluxes.

In the case of ASDEX-UG the main contribution to electron heat flux,
see figure \ref{heatfluxes} (b), is generated by the MTM instability
at low $k_{\theta}\rho_{s}$, and for higher $k_{\theta}\rho_{s}$ where
the most unstable mode switches to an ITG mode, the electron heat flux
is significantly reduced. {However, at the very high poloidal mode
  numbers $1< k_{\theta}\rho_{s}$ where an ETG is the dominant
  instability the electron heat flux increases again.} Also here,
$\delta A_{\parallel}$ (blue dash-dotted line) generates the dominant
contribution to the electron heat flux in the MTM dominated region
while $\delta \phi$ (red dashed line) produces the dominant
contribution to the electron heat flux in the {ITG/ETG dominated
  regions. For the ion heat flux the main contribution comes from the
  higher $k_{\theta}\rho_{s}$ region where the ITG mode is the most
  unstable mode present with the maximum around
  $k_{\theta}\rho_{s}\sim0.5$, as illustrated in figure
  \ref{heatfluxes} (e). The MTM and ETG contributions to the ion heat
  flux are significantly smaller. Also, for the electron and impurity
  particle fluxes, shown in figures \ref{particlefluxes} (b, e and h),
  the contributions from MTM and ETG is negligible compared to the
  contribution from the ITG.}

Similar trends are observed for heat and particle transport in the JET
case. Again here, in the MTM dominated region the electron heat flux,
see figure \ref{heatfluxes} (c), is the main channel of transport
while the ion heat flux, presented in figure \ref{heatfluxes} (f), and
the particle fluxes shown in figures \ref{particlefluxes} (c, f and i)
are negligibly small in comparison.

Remarkably, the main ion energy fluxes ($Q_i$) generated by the ITG
modes in the ASDEX-UG is found to be inward in spite of the positive
ion temperature gradient, see Figs.~\ref{heatfluxes} (d).  
However we
note, that in the ASDEX-UG case 1) there is a high inward particle
flux of main ions which might account for the inward energy flux if
most of it is convective, 2) the strong positive ion energy flux
carried by the nitrogen impurities almost cancel the negative energy
flux of the main ions, so the total ion energy flux is close to zero.

\subsection{Parametric dependences}
\begin{figure}[htbp]
\begin{center}
 \includegraphics[width=0.45\textwidth]{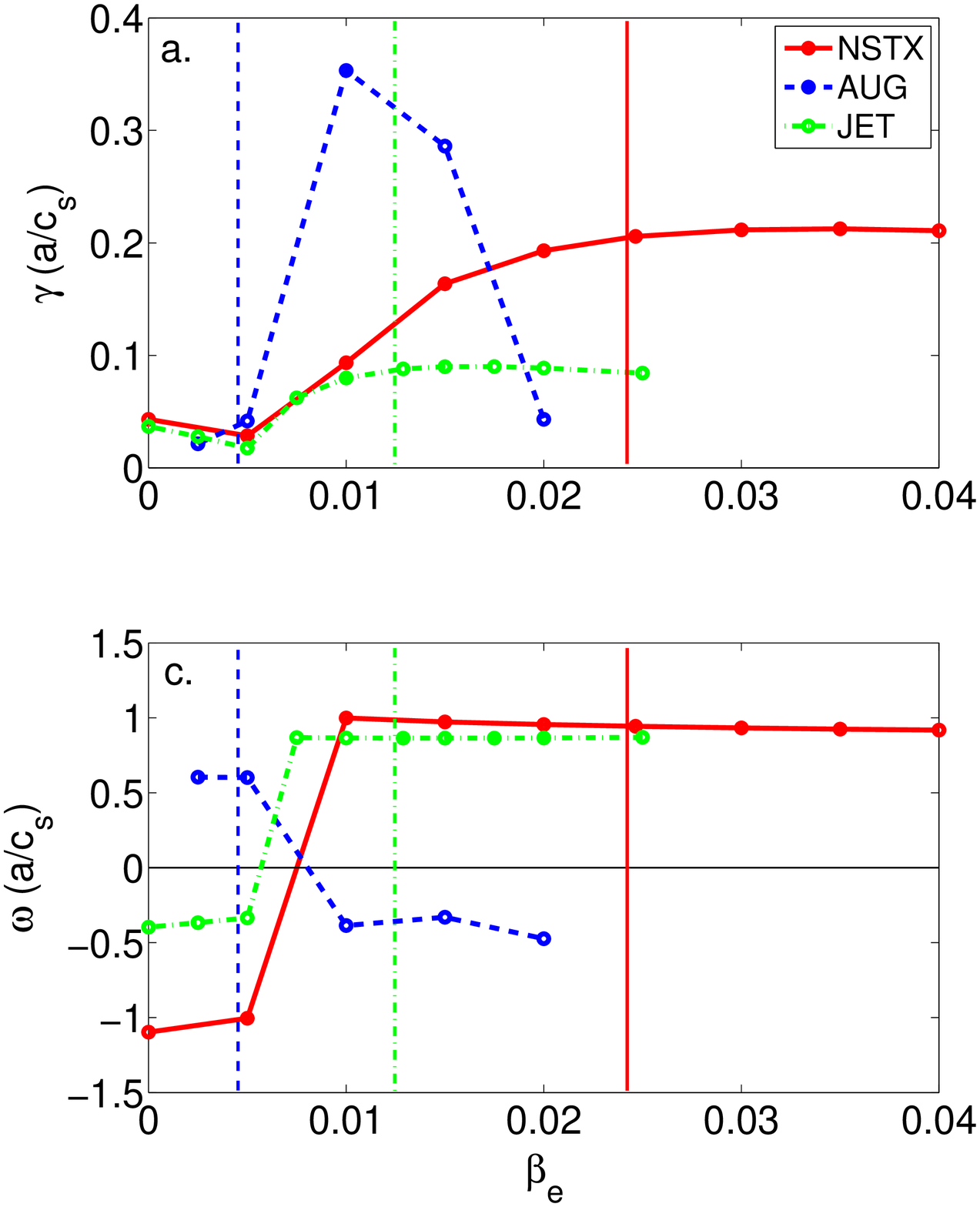} \includegraphics[width=0.45\textwidth]{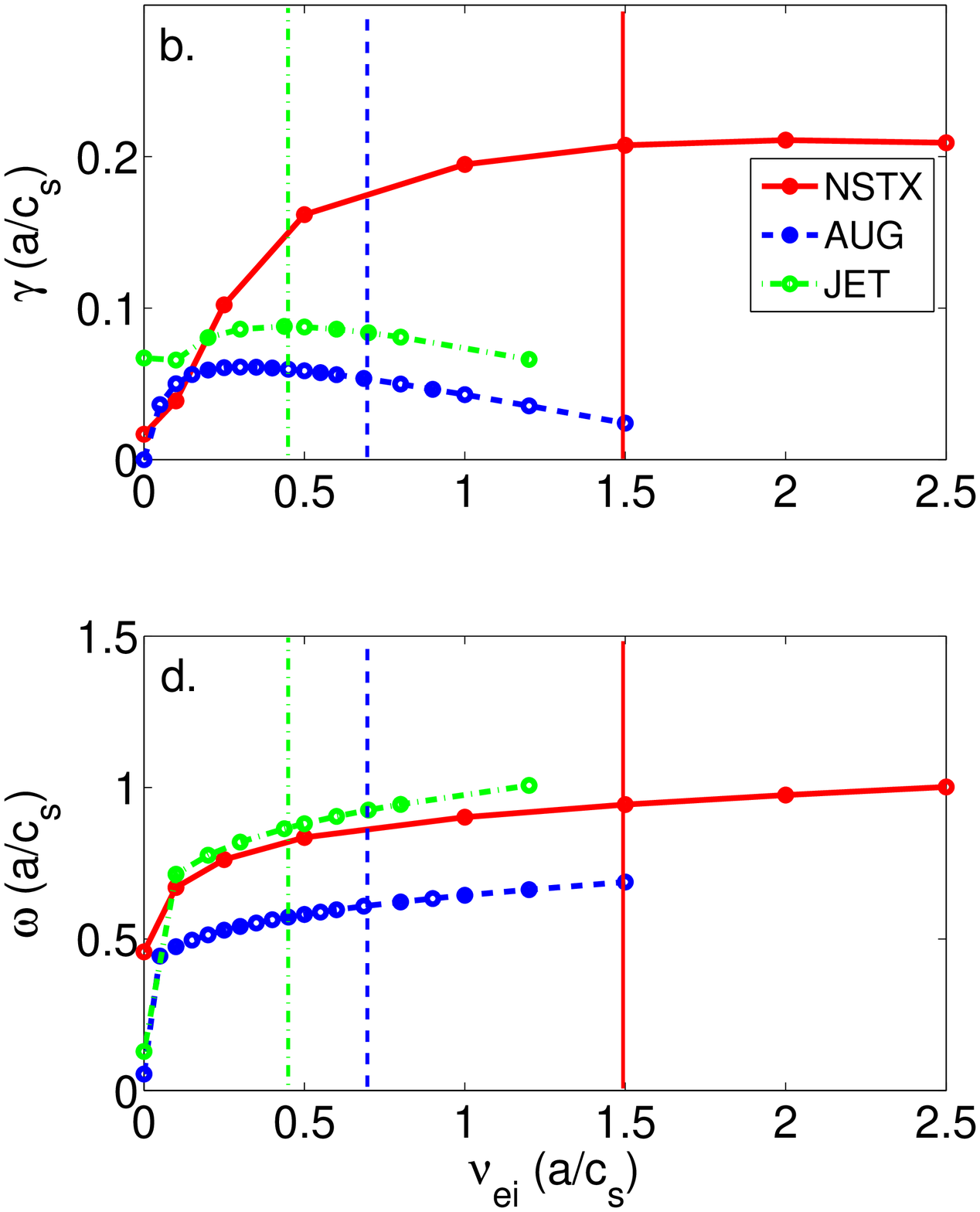}
 \caption{Imaginary and real parts of eigenvalues ($\gamma$,
   $\omega_{r}$) as functions of $\beta_e$ (a and
   c), and $\nu_{ei}$ (b and d).
   Red solid lines: NSTX, blue dashed: ASDEX-UG, green dash-dotted:
   JET. The vertical lines represent the base parameters color coded
   similarly to their respective machine.}
\label{betascan}
\end{center}
\end{figure}

Figures \ref{betascan} (a,c) show the linear growth rates and the real
frequencies of the {unstable modes, corresponding to
  the fixed values of $k_{\theta}\rho_{s}$ given in table
  \ref{table3}, as functions of $\beta_e$ for the three considered
  machines.} As seen in this figure for the experimental values of
$\beta_e$ (shown with vertical lines) the most unstable mode is an MTM
for all three tokamaks. In NSTX, the onset of MTM is well below the
experimental value of $\beta_e$ and an increase in $\beta_e$ above its
experimental value does not increase the growth rate
significantly. The same observation can be made for the JET case,
while for ASDEX-UG, although the MTM onset is well below its
experimental value, by further increasing $\beta_e$, a
Kinetic-Ballooning Mode (KBM) (negative real frequency in ion
direction) appears as the most unstable mode. These results show that
since MTMs become unstable due to electromagnetic perturbations a
finite level of $\beta_e$ is needed for their onset, but when
unstable, MTMs do not strongly depend on $\beta_e$ variations. In all
these scans the value of $\alpha_{MHD}$ is calculated consistently
with the local beta values and the density and temperature gradients.

\begin{figure}[htbp]
\begin{center}
 \includegraphics[width=0.32\textwidth]{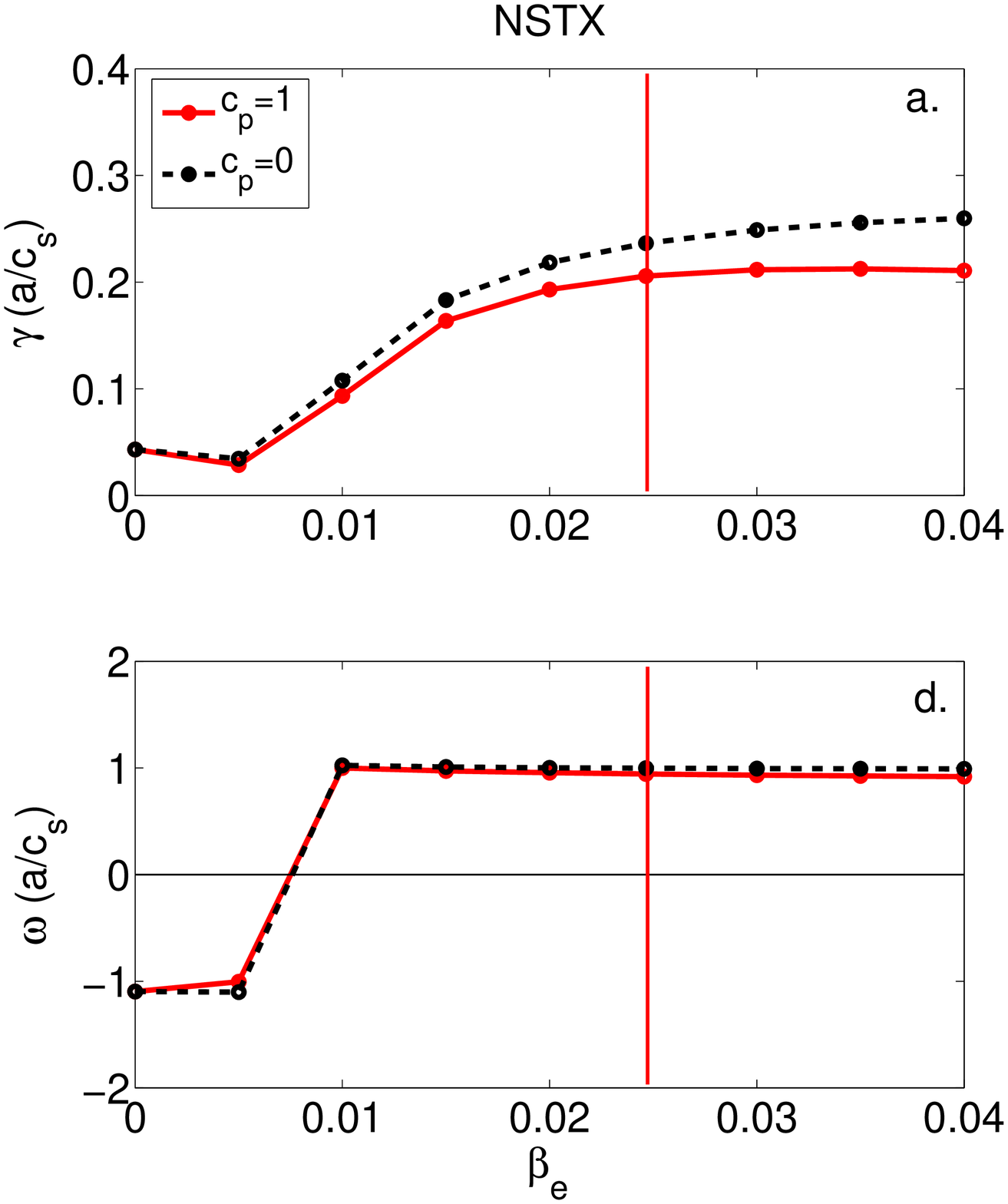} \includegraphics[width=0.32\textwidth]{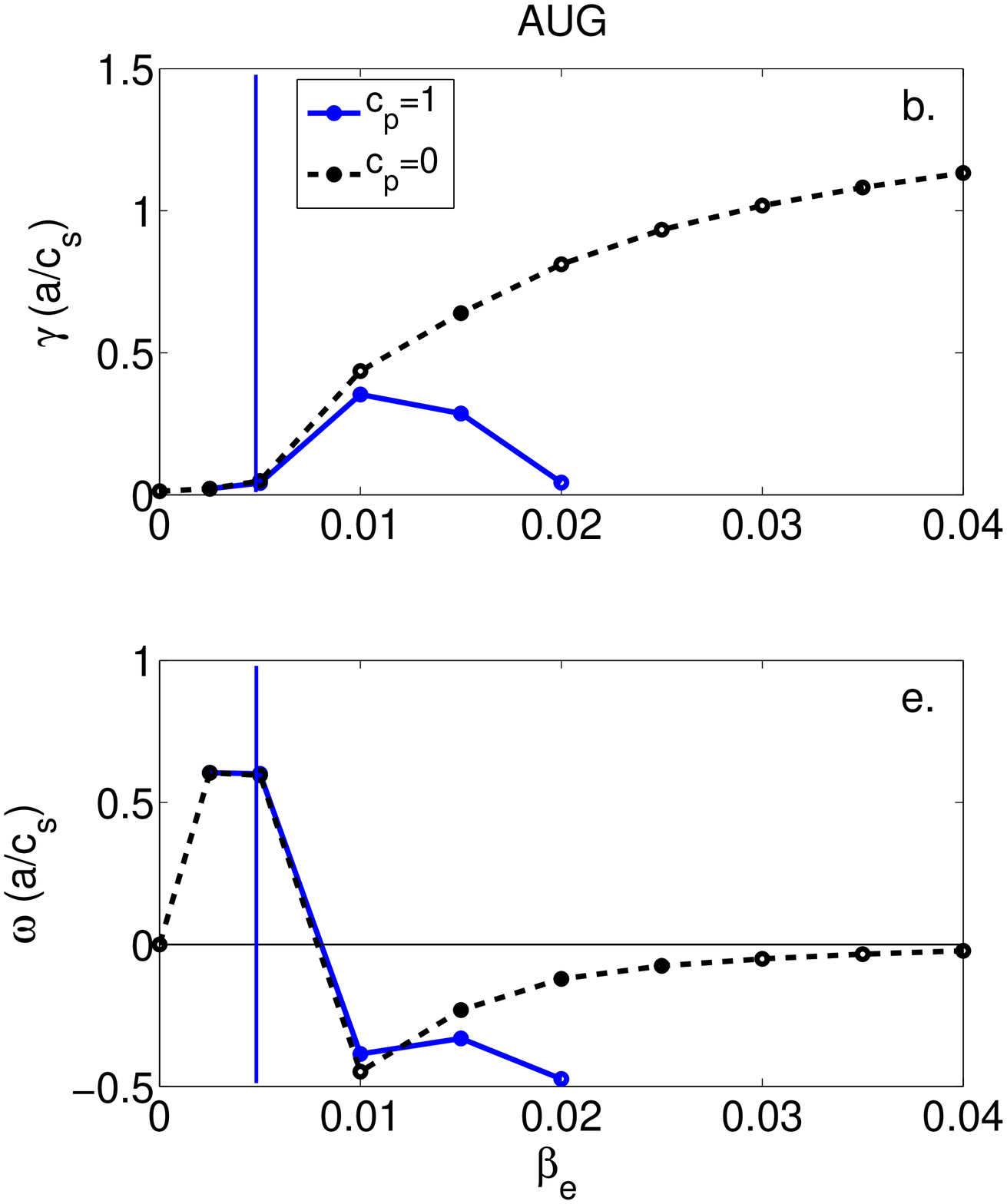} \includegraphics[width=0.32\textwidth]{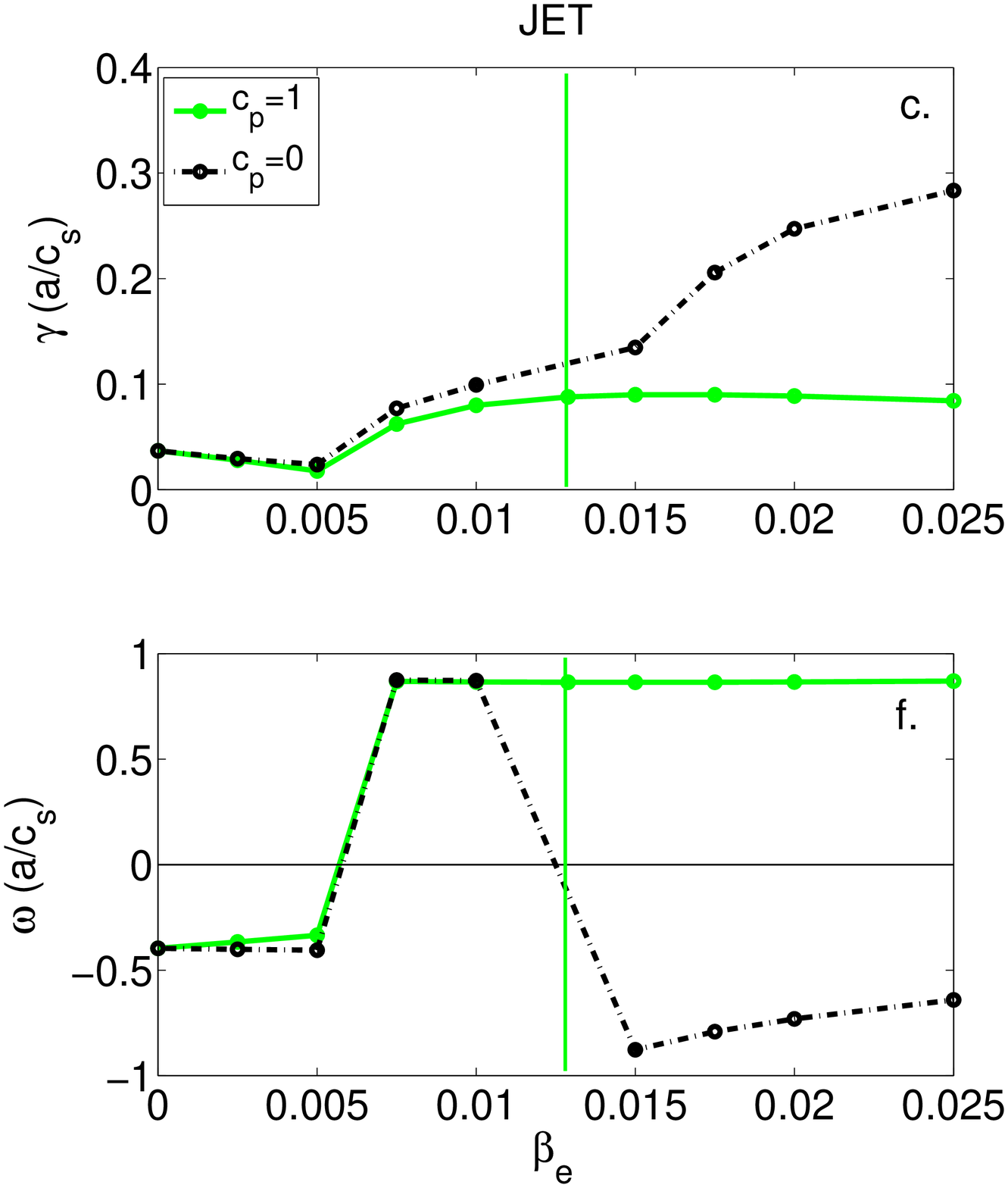}
 \caption{Imaginary and real parts of eigenvalues ($\gamma$,
   $\omega_{r}$) as functions of $\beta_e$ for $c_p=1$ (solid lines) and $c_p=0$ (dashed black lines).
     NSTX (a and d), ASDEX-UG (b and e), and JET (c and f). The vertical lines represent the base parameters color coded
   similarly to their respective machine.}
\label{betascan2}
\end{center}
\end{figure}

In order to determine the effect of $\alpha_{MHD}$ through the
curvature drift a $\beta_e$ scan, similar to figure \ref{betascan} (a
and c), is performed where $\alpha_{MHD}$ is scaled to zero by setting
$c_p=0$. The corresponding results are shown in figure
\ref{betascan2}. The $\alpha_{MHD}$-stabilization is not significant
in the MTM regime in any of the studied plasmas. In the NSTX case
there is a small $\alpha_{MHD}$-stabilization only at very high
$\beta_e$, see figures \ref{betascan2} (a,d). In the ASDEX-UG case,
the stabilization of the KBM mode at higher $\beta_e$ is clearly an
$\alpha_{MHD}$ effect, since without it the mode is further
destabilized by an increase in $\beta_e$, see the black dashed lines
in figures \ref{betascan2} (b,e). For JET, the situation is different,
as seen in figures \ref{betascan2} (c,f). In the electrostatic limit
and at low $\beta_e$ the most unstable mode is an ITG which is
stabilized as $\beta_e$ is increased even without $\alpha_{MHD}$
effect. For higher $\beta_e$ the dominant mode switches to an
MTM. Without $\alpha_{MHD}$ effects, by increasing $\beta_e$ even
further the MTM switches to a KBM that is further destabilized by
$\beta_e$, however with the $\alpha_{MHD}$ effect included the KBM
appears only at higher $\beta_e$ (outside the plotted $\beta_e$
range).

As the collisionality is suggested as an important parameter in
driving the MTMs unstable, here we examine the dependence of the MTM
mode characteristics on this parameter. In figures \ref{betascan}
(b,d) the linear growth rates and the real frequencies of the most
unstable modes are shown as functions of the collision frequency
$\nu_{ei}$ for the different tokamaks. For the experimental values of
$\nu_{ei}$ (marked with vertical lines) the most unstable mode is
found to be an MTM in all the three considered tokamaks. In NSTX, the
growth rate increases with collisionality as expected since
collisionality is one of the instability drives \cite{drack,catto},
however, a further increase in $\nu_{ei}$ above the experimental value
(almost doubled) the MTM growth rate does not increase significantly.
In the ASDEX-UG the growth rate shows a decline as collisionality
increases. This trend is not surprising since previous studies have
shown that the growth rate of the MTM has a non-monotonic dependence
on the collisionality
\cite{ApplegatePPCF2007,GuttenfelderPRL2011,DoerkPRL2011}. 
As the collisionality further increases beyond the peak
value, particles are very much scattered by collisions, and therefore
preventing the formation of the current layer necessary for the MTMs
to become unstable. Hence, as seen in figure \ref{betascan} (b) we
expect that this is the case for the ASDEX-UG case. For JET we also
observe a (gentle) non-monotonic trend, and our base value of the
collisionality seems to be positioned near the peak value. In all
three machines no unstable MTMs were found for the collisionless case,
i.e.~$\nu_{ei}=0$, and under this condition these plasmas are found to
be TEM unstable. Using the eigenvalue solver method in \gyro
\cite{gyro} we followed the root corresponding to the MTM instability
towards smaller $\nu_{ei}$, and the mode is completely stabilized in
the collisionless limit. The values of $\nu_{ei}/\omega_r$
corresponding to the maximum growth rate are rather different between
the spherical and the conventional tokamak cases; in NSTX it is
$2.05$, while it is $0.55$ and $0.51$ in the AUG and the JET cases,
respectively.

\begin{figure}[htbp]
\begin{center}
 \includegraphics[width=0.45\textwidth]{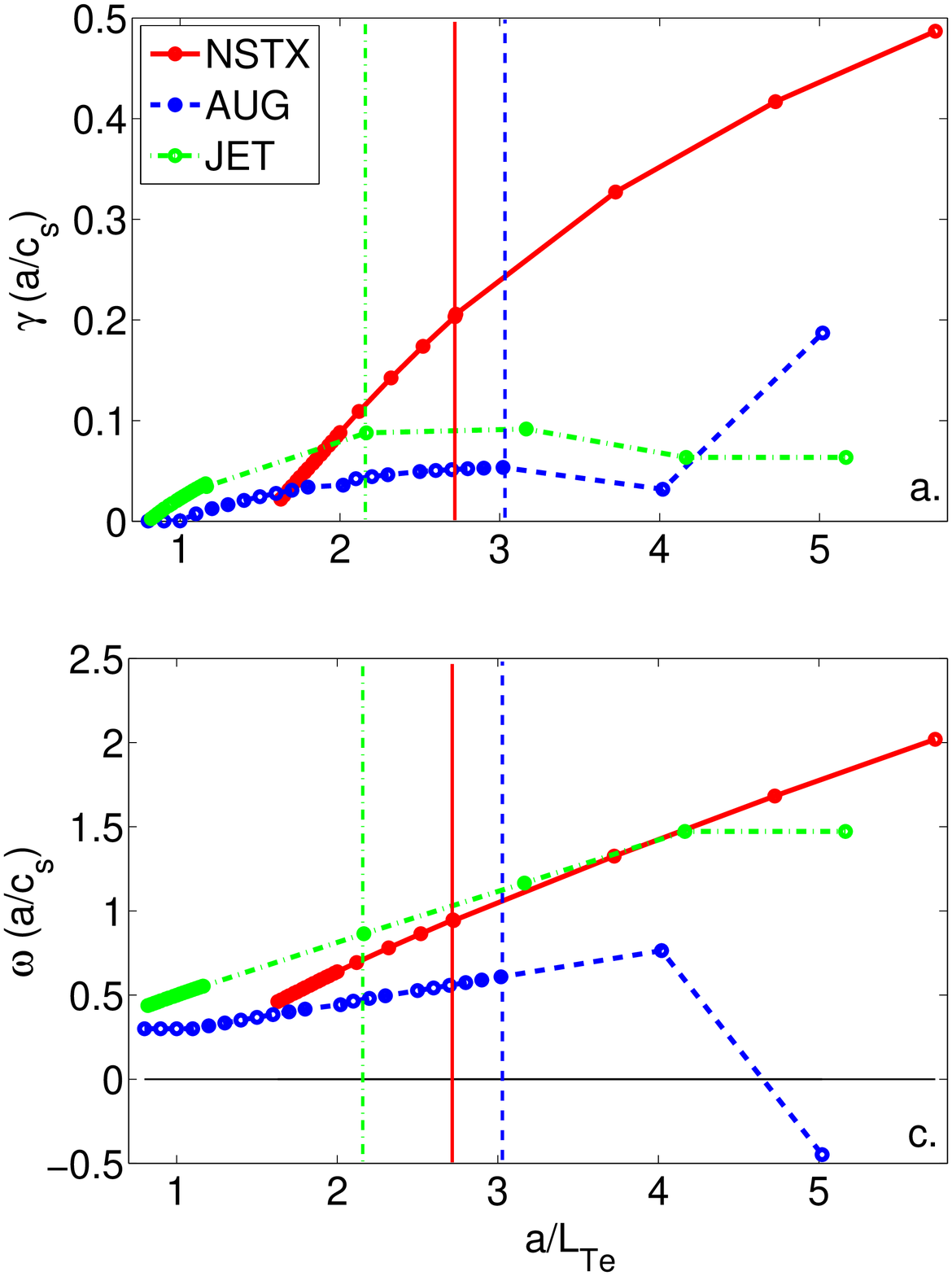}  \includegraphics[width=0.45\textwidth]{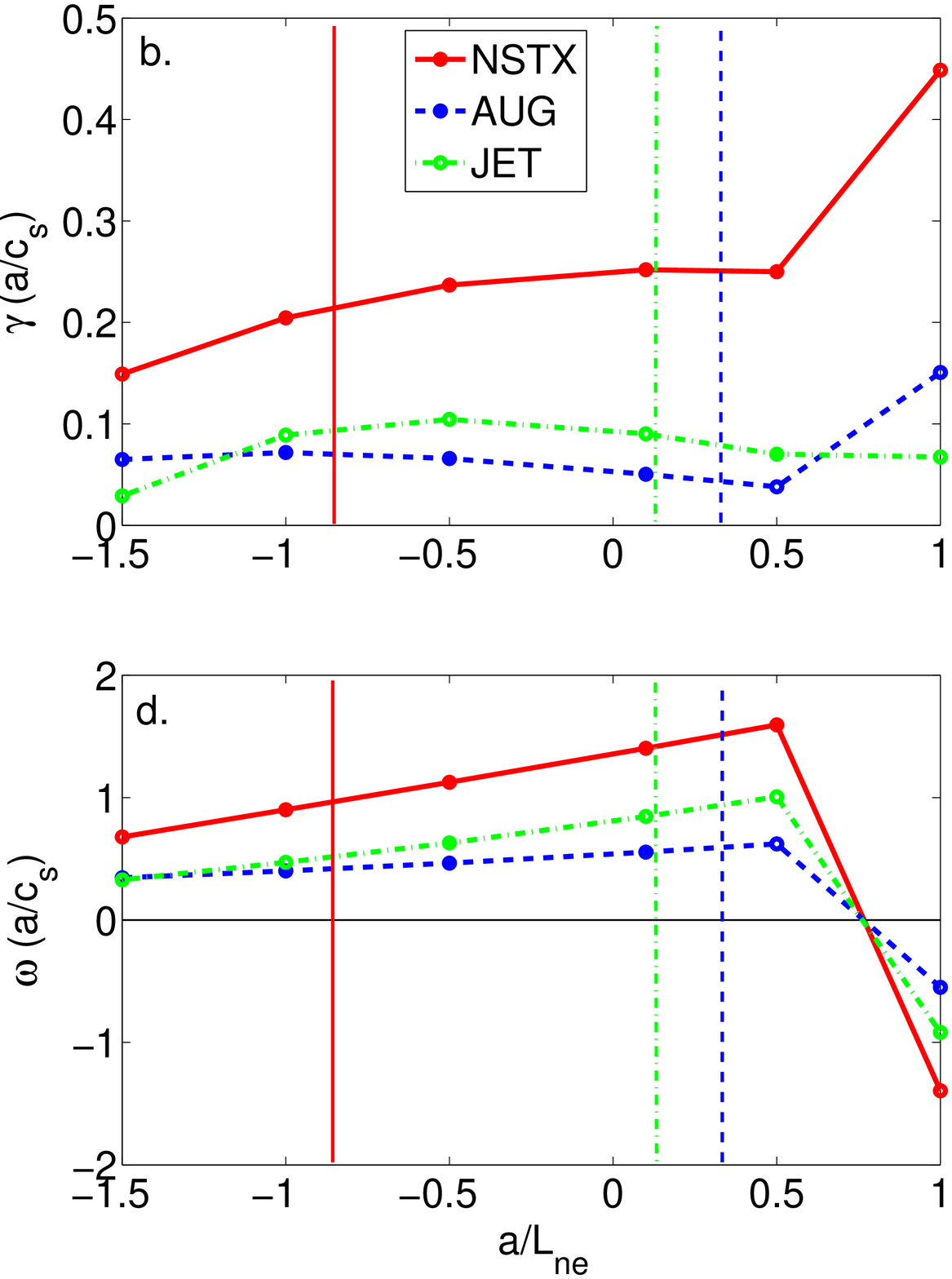}
 \caption{Imaginary and real parts of eigenvalues ($\gamma$,
   $\omega_{r}$) as functions of $a/L_{Te}$ (a and c), and
   $a/L_{ne}$ (b and d). Red solid lines: NSTX, blue dashed: ASDEX-UG,
   green dash-dotted: JET. The vertical lines represent the base
   parameters color coded similarly to their respective
   machine.}
\label{altescan}
\end{center}
\end{figure}

In the literature the electron temperature gradient is suggested as
one of the instability drives for MTMs \cite{drack,catto}. This has
been confirmed in previous numerical studies
\cite{ApplegatePPCF2007,GuttenfelderPRL2011,DoerkPRL2011}. Here we
compare the role of electron temperature gradient in destabilization
of the MTMs between the three considered machines by performing a scan
over the $a/L_{Te}$ parameter. The results of this scan are shown in
figures \ref{altescan} (a,c) where the linear growth rates and the
real frequencies of the most unstable modes are illustrated as
functions of $a/L_{Te}$. As seen in this figure, a finite value of
$a/L_{Te}$ is necessary for the destabilization of the MTMs in all
three machines, however a clear difference is observed in the
variation of MTM growth rates with $a/L_{Te}$ between spherical and
conventional tokamaks. For the NSTX case, there is a clear increase of
the MTM growth rate with an increase in $a/L_{Te}$, while for ASDEX-UG
and JET the MTM growth rates show weaker and a non-monotonic
dependence on $a/L_{Te}$. By further increase in $a/L_{Te}$ the most
unstable mode switches from an MTM to ITG/TEM modes, corresponding to
the last points in the ASDEX-UG and JET curves in Figs.~\ref{altescan}
(a and c). 

\begin{figure}[htbp]
\begin{center}
 \includegraphics[width=0.32\textwidth]{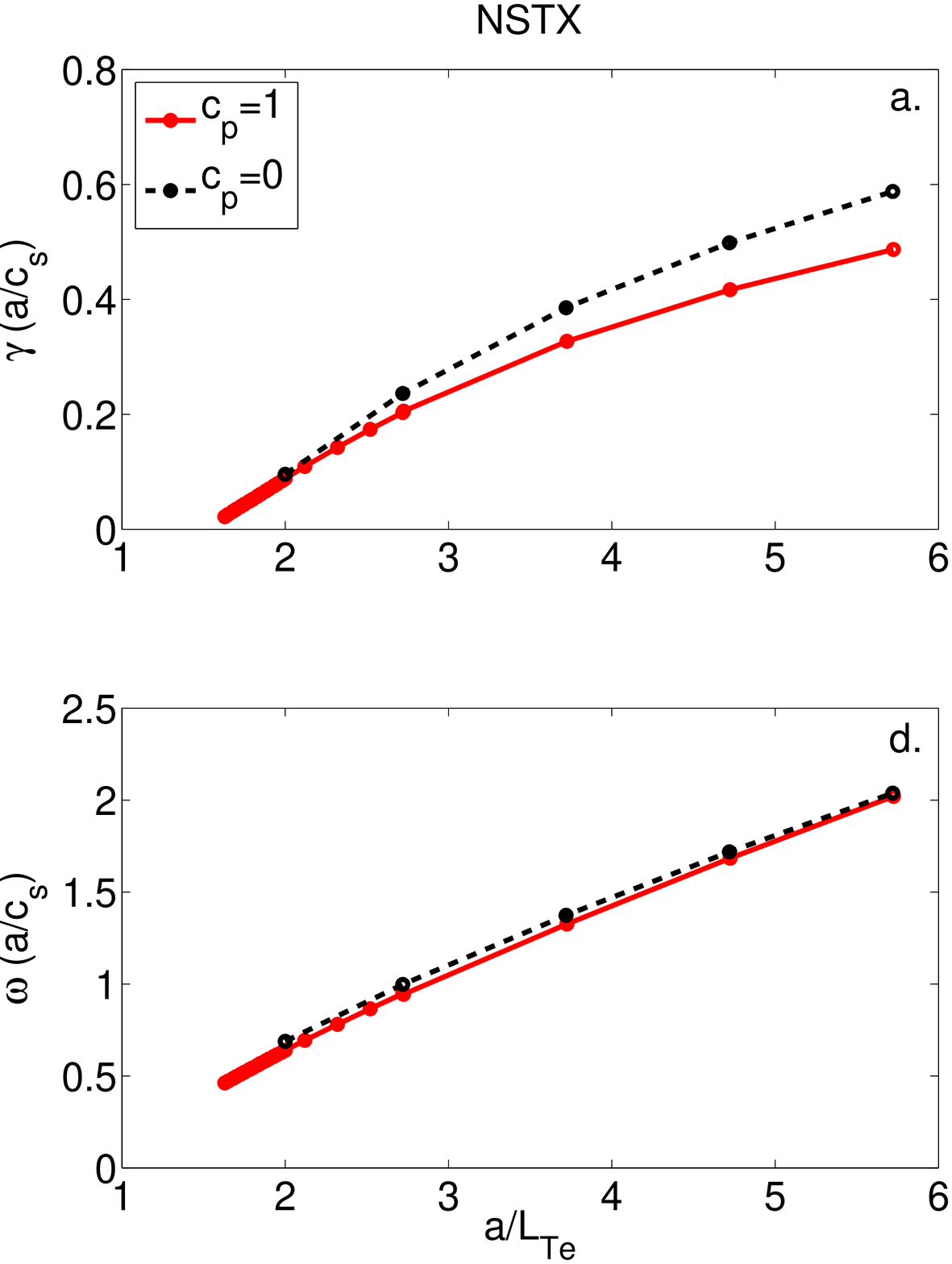} \includegraphics[width=0.32\textwidth]{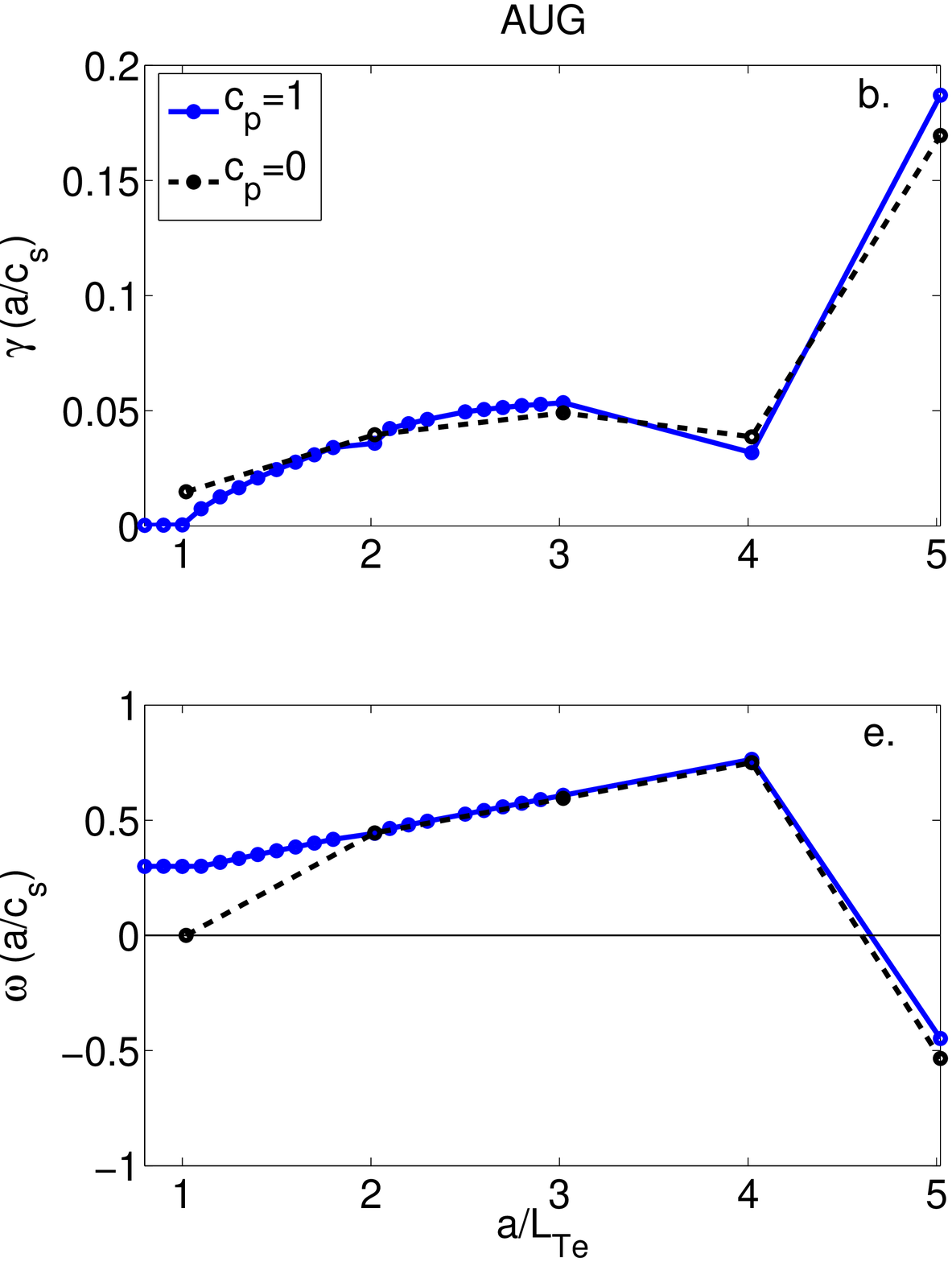} \includegraphics[width=0.32\textwidth]{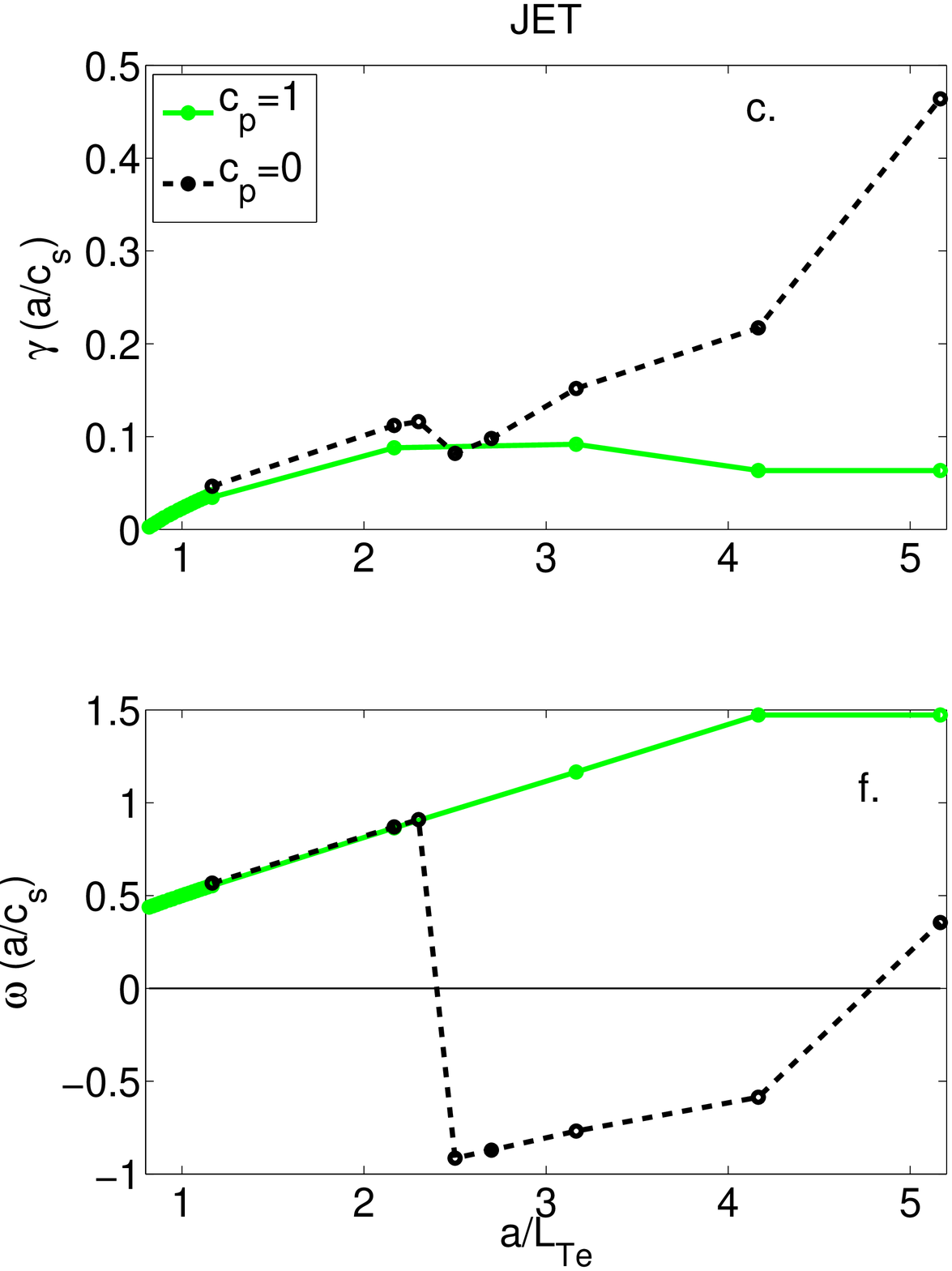}
 \caption{Imaginary and real parts of eigenvalues ($\gamma$,
   $\omega_{r}$) as functions of $a/L_{Te}$ with $c_p=1$ and $c_p=0$. Red solid lines: NSTX, blue dashed: ASDEX-UG,
   green dash-dotted: JET.}
\label{altewoalpha}
\end{center}
\end{figure}

The $a/L_{Te}$ threshold for MTM instability is observed to be well
below our baseline values (indicated by vertical lines), and it is
lower for JET and ASDEX-UG than for NSTX.  The growth rate of MTMs is
found to be less sensitive to electron temperature gradient for both
of the conventional tokamaks than in the spherical tokamak studied
here. In order to investigate the reason for this difference we
performed a similar scan over $a/L_{Te}$ and set $c_p=0$ to eliminate
the $\alpha_{MHD}$ stabilization effect. The results are shown in
figure \ref{altewoalpha}. Without $\alpha_{MHD}$ stabilization the MTM
still remains the most unstable mode for the baseline parameters, and
as $a/L_{Te}$ increases no significant change is observed for NSTX and
ASDEX-UG. However, for JET case three regions in the $a/L_{Te}$ space
can be distinguished.  For low values of $a/L_{Te}$ the MTM is the
dominant instability; its growth rate increases linearly with
$a/L_{Te}$. At $a/L_{Te}\sim 2.5$ an ITG mode, which was previously
stabilized by $\alpha_{MHD}$ effects -- see figures \ref{altescan}
(a,c), -- becomes the dominant instability, which gradually transits
to a TEM as $a/L_{Te}$ becomes very large. Therefore, from our
observations the stabilizing effect of the $\alpha_{MHD}$ parameter on
the MTM mode is not significant, but it has an impact on the stability
of ITG/TEM{/KBM modes.  The strong suppression of these modes
  therefore, allows for MTMs to remain the dominant instability} for a
wider range of $a/L_{Te}$. The weaker and non-monotonic dependence of
the MTM on the electron temperature gradient in ASDEX-UG and JET thus,
can not be explained by this effect.

One of the main differences between the plasma parameters in the three
plasmas considered appears in the $a/L_{ne}$ values; a strongly
negative electron density gradient, corresponding to a hollow electron
density profile, is observed in NSTX, while the density profiles were
slightly peaked in the ASDEX-UG and JET plasmas. Thus, we have
investigated the dependence of MTM linear growth rates on the
$a/L_{ne}$ parameter. Quasi-neutrality is enforced by slightly varying
$a/L_{ni}$ while keeping the impurity density gradients fixed to the
base parameters given in table \ref{table2}. Figures \ref{altescan}
(b-d) show the linear growth rates and the
real frequencies of the most unstable modes versus $a/L_{ne}$ for the
different machines. In all three machines, the MTMs linear growth
rates exhibit a non-monotonic dependence on $a/L_{ne}$ parameter with
maxima corresponding to slightly hollow electron density profiles
($a/L_{ne}\sim-0.5$). Clearly $a/L_{ne}$ is not a strong and necessary
drive for the MTMs, as there are finite MTM growth rates in all
machines at $a/L_{ne}=0$.

\begin{figure}[htbp]
\begin{center}
 \includegraphics[width=0.32\textwidth]{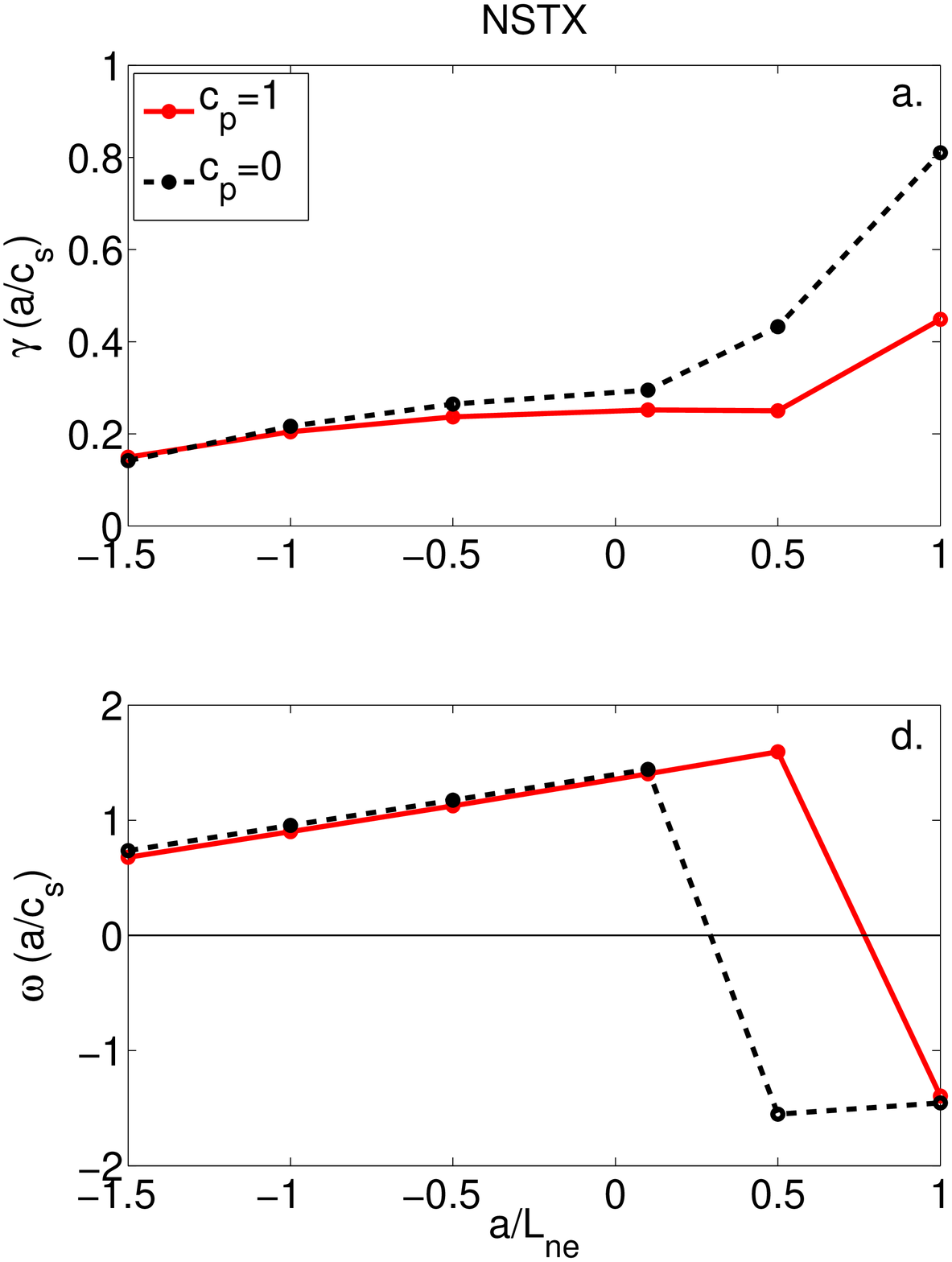} \includegraphics[width=0.32\textwidth]{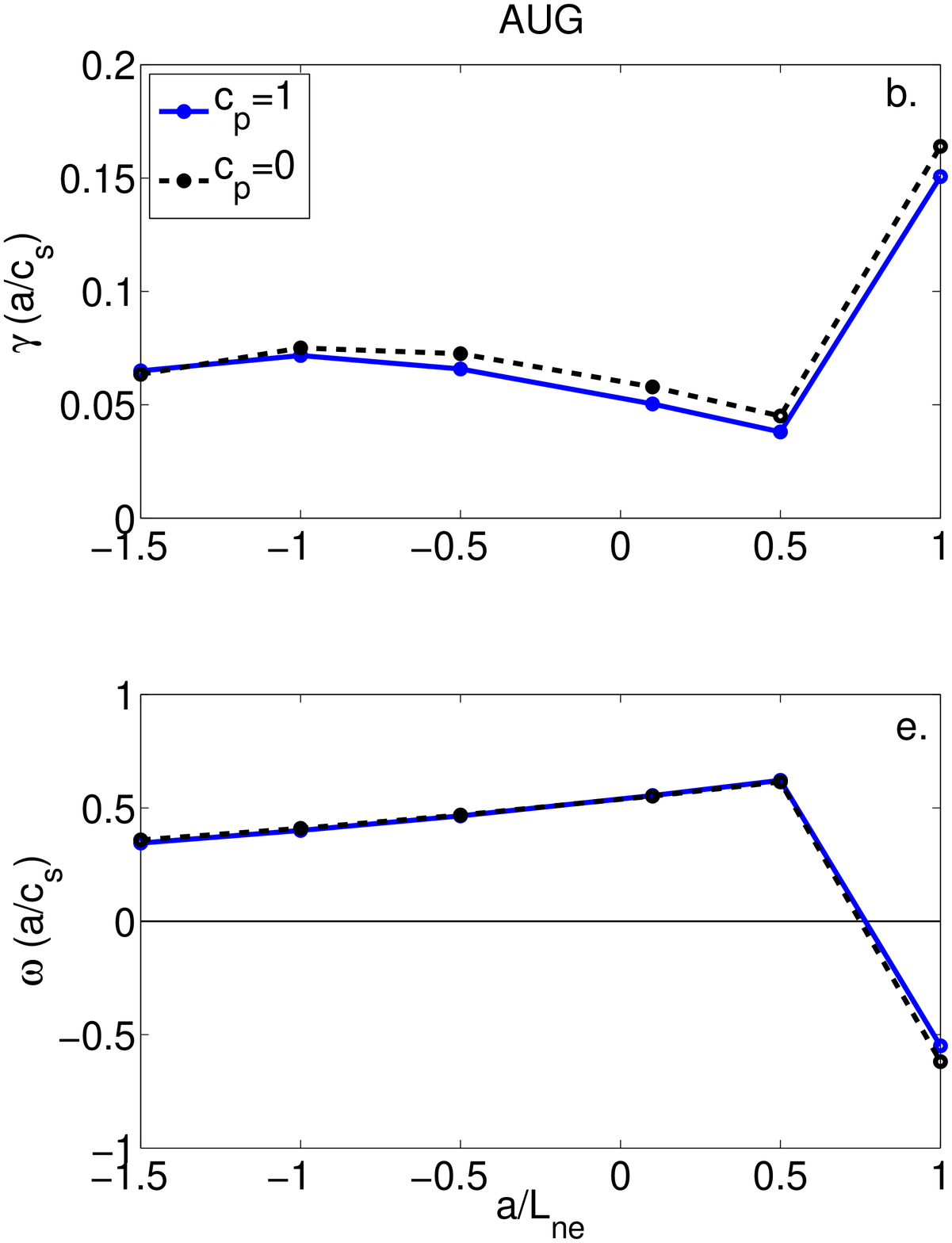} \includegraphics[width=0.32\textwidth]{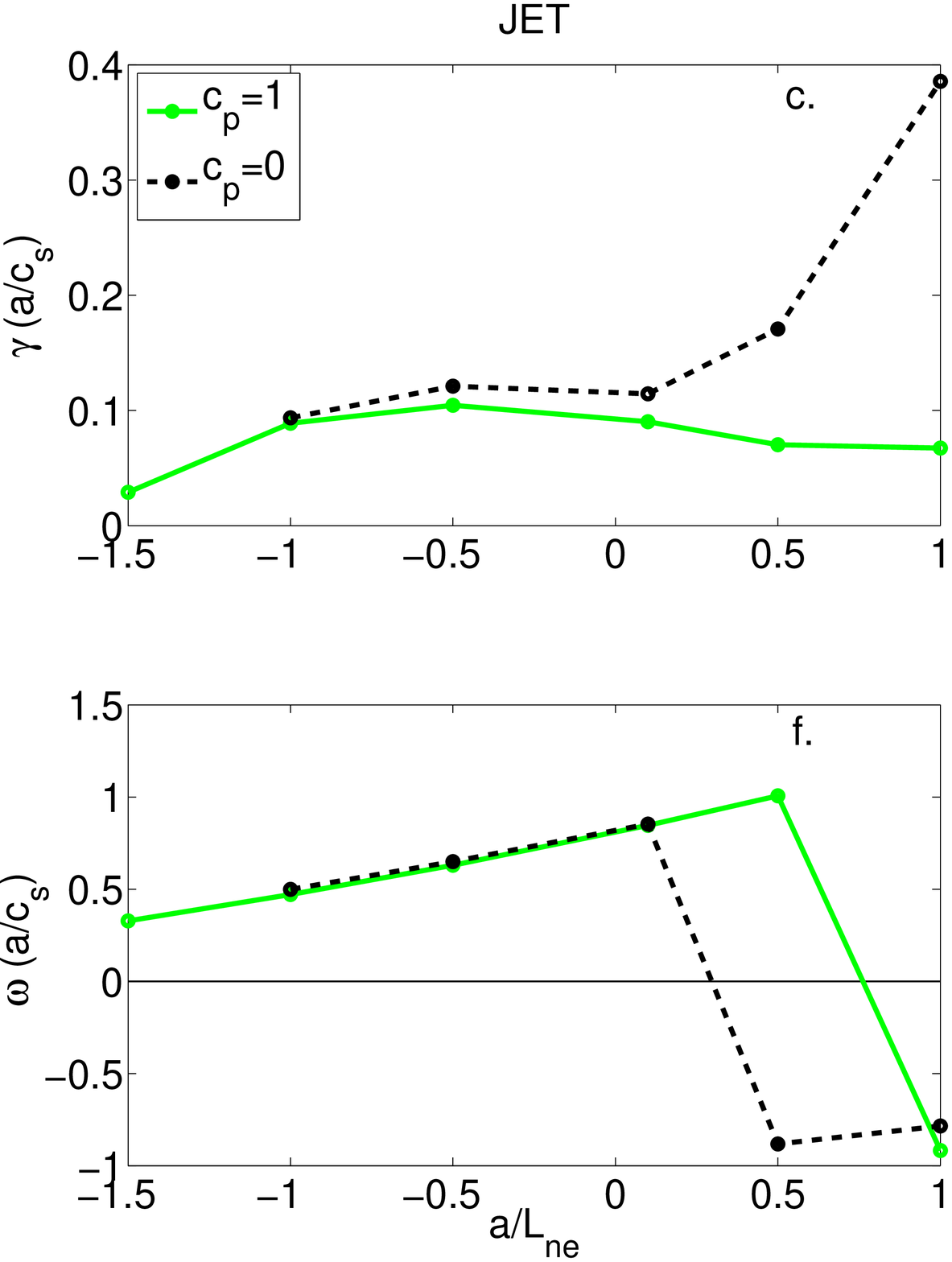}
 \caption{Imaginary and real parts of eigenvalues ($\gamma$,
   $\omega_{r}$) as functions of $a/L_{ne}$ with $c_p=1$ and $c_p=0$. Red solid lines: NSTX, blue dashed: ASDEX-UG,
   green dash-dotted: JET.}
\label{alnewoalpha}
\end{center}
\end{figure}

In all the machines the MTM is the most unstable mode over the range
of $-1.5\le a/L_{ne} \le 0.5$, with a rather weak dependence on this
parameter. For sufficiently high electron density gradient the
dominant linear mode switches from a MTM to an ITG mode in all
plasmas. Again we examined the impact of $\alpha_{MHD}$ by comparing
the results to corresponding simulations with $c_p=0$ (black dashed
lines in figure \ref{alnewoalpha}). Similar to the results of the
$a/L_{Te}$ scans shown in figure \ref{altewoalpha}, the effect of the
$\alpha_{MHD}$ parameter on the MTM mode itself is negligible, but it
strongly stabilizes the ITG mode. As shown in figure
\ref{alnewoalpha}, without $\alpha_{MHD}$ effects in the cases of NSTX
and JET the $a/L_{ne}$ threshold of the dominant ITG mode is reduced,
but in ASDEX-UG case no significant change is observed. 


\section{Conclusions}
\label{sec:conclusions}
We have investigated the onset and parametric dependences of the MTM
instability in the core ($\rho_{tor}=0.6$) of a spherical (NSTX), and
two conventional tokamaks (ASDEX-UG, JET). The quasilinear transport
is computed using the gyrokinetic code \gyro in the flux-tube (local)
limit. In confirmation with previous studies, we found that for the
experimentally relevant plasma parameters the MTMs are linearly the
dominant instability in NSTX and ASDEX-UG. Under typical JET baseline
parameters considered here the MTMs are also found as the dominant
linear instability. In NSTX and JET the maximum of the MTM linear
growth rate is located at higher mode numbers than that for
ASDEX-UG. Therefore, the previously discussed idea that the higher
mode number MTMs are the characteristics of the spherical tokamaks,
while lower mode number MTMs are the characteristics of the
conventional tokamaks is not supported by our results.

Parametric scaling of the MTM instability in the core of these
plasmas, revealed that a finite level of $\nu_{ei}$ and $\beta_e$ are
needed in order for MTMs to become unstable, and when unstable, they
can remain the dominant instability over a wide range in $\nu_{ei}$ and
$\beta_e$. The linear MTM growth rate seems to be only weakly
dependent on $\beta_e$, and exhibits a weak but non-monotonic 
dependence on collisionality, $\nu_{ei}$. By neglecting the collisionality or
$\beta_e$ effects, the ITG/TEM modes appear as the most unstable modes
in all three studied machines.

A strong dependence for the growth rate of MTMs on electron
temperature gradient is found in NSTX, while for ASDEX-UG and JET the
MTM growth rate is found to be less sensitive to this parameter. The
MTM growth rate significantly increases as $a/L_{Te}$ increases in
NSTX, but in ASDEX-UG and JET a weak and non-monotonic dependence on
$a/L_{Te}$ is found. These results indicate that while $a/L_{Te}$ is
a fundamental drive for the MTMs in these plasmas, it can contribute
to the stabilization of the mode as well; this non-trivial behavior is
more pronounced in the ASDEX-UG and JET plasmas.

Similar trends are observed in all three machines when scanning for
the electron density gradient, where the MTMs linear growth rates
show again a non-monotonic dependences on $a/L_{ne}$ parameter with peaks
located in the negative $a/L_{ne}$ region corresponding to slightly
hollow electron density profiles.

We have investigated the impact of a finite $\alpha_{MHD}$ on the
onset and characteristics of MTM instability in various parametric
scans, and we have observed that the stabilization of $\alpha_{MHD}$
parameter on the MTM mode itself is not significant in all
machines. However, its impact on the ITG/TEM unstable modes can result
in a strong suppression of these modes allowing for the MTM to remain
the dominant instability for a wider range in considered parameters,
i.e. $\beta_e$, $a/L_{Te}$ and $a/L_{ne}$. The non-monotonic and
weaker dependence of the MTM on the electron temperature/density
gradient however, can not be explained by $\alpha_{MHD}$ effects.

In the studied cases the MTM drives mostly electron heat transport
through $\delta A_{\parallel}$ fluctuations, while other transport
channels and contributions from $\delta \phi$ and $\delta
B_{\parallel}$ are significantly smaller. For the ion heat flux and
particle fluxes, the transport driven by MTMs are negligible, however,
if ballooning modes like ITG/TEM/KBM are also present, even as sub-dominant
modes, these fluxes are mainly driven by these modes and therefore the
overall ion heat and particle transport may not be
negligible. However, these findings are based on linear analysis and
further non-linear studies are needed for their confirmation.

Finally, we would like to stress that the cases which we have
considered have been chosen because they were unstable to MTMs.
However, particularly large aspect ratio tokamaks, being unstable to
MTMs should not be considered a generic property.

\section*{Acknowledgments}
The authors would like to thank F Jenko, J Candy and C Angioni for
valuable comments, and J Candy for providing the \gyro code. This work
was funded by the European Communities under Association Contract
between EURATOM and {\em Vetenskapsr{\aa}det}. The views and opinions
expressed herein do not necessarily reflect those of the European
Commission.

\end{document}